\documentclass[fleqn,usenatbib]{mnras}

\usepackage{newtxtext,newtxmath}
\usepackage[T1]{fontenc}

\DeclareRobustCommand{\VAN}[3]{#2}
\let\VANthebibliography\thebibliography
\def\thebibliography{\DeclareRobustCommand{\VAN}[3]{##3}\VANthebibliography}

\usepackage{graphicx}
\usepackage{xcolor}
\usepackage{amsmath}
\usepackage{soul}
\usepackage{array}
\usepackage{bm}
\usepackage{verbatim}
\graphicspath{{./figures/}{../figures/}}
\usepackage{lastpage}

\newcommand{\ts}{\textsection}
\newcommand{\tn}[1]{\textnormal{#1}}
\newcommand{\expect}[1]{\left\langle#1\right\rangle}

\newcommand{\solmass}{\tn{M}_{\odot}}
\newcommand{\Mstar}{\tn{M}_{*}}
\newcommand{\Mbh}{\tn{M}_{\tn{BH}}}

\newcommand{\Lbol}{\tn{L}_{\tn{bol}}}
\newcommand{\Mbulge}{\tn{M}_{*\tn{bulge}}}

\newcommand{\MdotBH}{\dot{\tn{M}}_{\tn{BH}}}
\newcommand{\MdotStar}{\dot{\tn{M}}_*}

\newcommand{\preslope}{\alpha}
\newcommand{\postnorm}{\beta}
\newcommand{\MstarCRIT}{\tn{M}_{*\tn{crit}}}
\newcommand{\MbhCRIT}{\tn{M}_{\tn{BHcrit}}}
\newcommand{\MbhNORM}{\gamma}

\newcommand{\Mbhpre}{\tn{M}_{\tn{BHlo}}}
\newcommand{\Mbhpost}{\tn{M}_{\tn{BHhi}}}
\newcommand{\siglnX}{\sigma_{\ln\tn{X}}}
\newcommand{\siglnXpre}{\sigma_{\ln\tn{Xlo}}}
\newcommand{\siglnXpost}{\sigma_{\ln\tn{Xhi}}}
\newcommand{\mulnX}{\mu_{\ln\tn{X}}}
\newcommand{\siglnMdot}{\sigma_{\ln\MdotBH}}
\newcommand{\mulnMdot}{\mu_{\ln\MdotBH}}

\title[Delayed SMBH Growth vs. the QLF]{Running Late: Testing Delayed Supermassive Black Hole Growth Models Against the Quasar Luminosity Function}

\author[Tillman et al.]{Megan Taylor Tillman,$^{1,2,3}$\thanks{E-mail: mtt74@physics.rutgers.edu}
Sarah Wellons,$^{2}$
Claude-Andr\'e Faucher-Gigu\`ere$^{2}$,
Luke Zoltan Kelley,$^{2}$
 \newauthor 
and Daniel Angl\'es-Alc\'azar$^{4,5}$
\\
$^{1}$Department of Physics and Astronomy, Texas A\&M University, College Station, TX 77843, USA\\
$^{2}$Center for Interdisciplinary Exploration and Research in Astrophysics (CIERA) \& Department of Physics and Astronomy, Northwestern University,\\ Evanston, IL 60201, USA\\
$^{3}$Department of Physics \& Astronomy, Rutgers, The State University of New Jersey, 136 Frelinghuysen Rd, Piscataway, NJ 08854, USA \\
$^{4}$Department of Physics, University of Connecticut, 196 Auditorium Road, U-3046, Storrs, CT 06269-3046, USA\\
$^{5}$Center for Computational Astrophysics, Flatiron Institute, New York, NY 10010, USA
}

\date{Accepted XXX. Received YYY; in original form ZZZ}

\pubyear{2022}

\begin{document}
\label{firstpage}
\pagerange{\pageref{firstpage}--\pageref{lastpage}}
\maketitle

\begin{abstract}
Observations of massive galaxies at low redshift have revealed approximately linear scaling relations between the mass of a supermassive black hole (SMBH) and properties of its host galaxy. How these scaling relations evolve with redshift and whether they extend to lower-mass galaxies however remain open questions. Recent galaxy formation simulations predict a delayed, or ``two-phase,'' growth of SMBHs: slow, highly intermittent BH growth due to repeated gas ejection by stellar feedback in low-mass galaxies, followed by more sustained gas accretion that eventually brings BHs onto the local scaling relations. The predicted two-phase growth implies a steep increase, or ``kink,'' in BH-galaxy scaling relations at a stellar mass $\Mstar\sim 5\times10^{10}$ M$_{\odot}$. We develop a parametric, semi-analytic model to compare different SMBH growth models against observations of the quasar luminosity function (QLF) at $z\sim0.5-4$. We compare models in which the relation between SMBH mass and galaxy mass is purely linear versus two-phase models. The models are anchored to the observed galaxy stellar mass function, and the BH mass functions at different redshifts are consistently connected by the accretion rates contributing to the QLF. The best fits suggest that two-phase evolution is significantly preferred by the QLF data over a purely linear scaling relation. Moreover, when the model parameters are left free, the two-phase model fits imply a transition mass consistent with that predicted by simulations. Our analysis motivates further observational tests, including measurements of BH masses and AGN activity at the low-mass end, which could more directly test two-phase SMBH growth.
\end{abstract}


\section{Introduction}
The co-evolution of supermassive black holes (SMBHs) and their host galaxies has been an active area of research for over two decades, driven in large part by observations \citep[for a comprehensive review, see][]{kormendy&ho2013}. 
Indeed, studies made possible by the Hubble Space Telescope revealed that all (or nearly all) low-redshift, massive galaxies host a nuclear SMBH. 
Moreover, detailed observations have shown that masses of SMBHs, $\Mbh$, correlate surprisingly tightly with properties of their host galaxies, such as the stellar bulge mass $\Mbulge$ \citep[e.g.,][]{Magorrian1998,haring+2004,marconi+2003}, or the velocity dispersion $\sigma$ of the stellar bulge \citep[e.g.,][]{Ferrarese2000,Gebhardt2000,Tremaine2002}.

    The local $\Mbh$-$\Mbulge$ scaling relation has generally been found to be nearly linear.
    However, this relation is largely constrained by measurements of relatively high mass galaxies in the local universe. 
    The empirical constraints on the $\Mbh$-$\Mbulge$ relation remain comparatively poor at lower masses and high redshift, despite a number of important observational efforts. 
    High-redshift measurements rely on indirect methods to estimate BH masses, such as emission line widths in active galactic nuclei \citep[AGN; e.g.,][]{Treu2007, Merloni2010, Shen2015}, while more direct measurements in low-redshift dwarf galaxies are  limited by sample sizes \citep[e.g.,][]{Lasker2016, Nguyen2019, Schutte2019}.
    Recent studies have also noted that the scaling relation appears to depend on the sample selection, for example early- vs. late-type or blue vs. red galaxies \citep{Graham+2013, Savorgnan+2016, Sahu+2019}. 
    These observational results imply that extrapolating the usual, linear scaling relation down to lower-mass galaxies may not be correct in general.
    
    Constraining the form of BH-galaxy scaling relations across the entire spectrum of galaxy masses and types, including in the early universe, is important not only for its own sake but also because it would allow a better understanding of how SMBHs grow with and affect their host galaxies via AGN feedback. 
    AGN feedback is a key ingredient in modern galaxy formation theories,
    but how exactly SMBHs couple to their host galaxies and halos remains a major unknown \citep[e.g.,][]{SD2015_ARAA, NO2017_ARAA}. 
    In current models, AGN feedback is usually assumed to be critical for quenching star formation in massive galaxies, which is needed to explain the observed sequence of ``red and dead'' galaxies \citep[e.g.,][]{Faber2007, Hopkins2008_red_ellipticals, Chen2020}. 
    However, there is increasing observational evidence of AGN-driven outflows in dwarf galaxies \citep[e.g.,][]{ManzanoKing2019,Liu2020}.
    This suggests that AGN feedback could be important for lower-mass galaxies as well. 
    AGN feedback can in principle affect BH-galaxy scaling relations either through its effect on star formation in the host galaxy or by regulating the growth of the nuclear BH. 

    This paper focuses on testing a prediction from a number of recent galaxy formation simulations concerning the growth of SMBHs. 
    In the last several years, cosmological simulations of galaxy formation have advanced greatly
    both in resolution (especially in ``zoom-in'' simulations) as well as in how small-scale processes such as star formation, stellar feedback, and black hole physics are modeled \citep[for reviews of recent progress, see][]{FG18_NA, Vogelsberger2020_NatRP}.  
    Although simulations using different codes differ in many important details, some predictions appear generic to relatively wide variations in simulation methodologies. 
    A well-known example of this is the role of stellar feedback in shaping the low-mass end of the galaxy stellar mass function \citep[e.g.,][]{SD2015_ARAA}. 
    Here, we are motivated by another prediction which appears generic to many different simulations, namely the delayed growth of SMBHs, with respect to stellar mass growth, due to gas ejection by stellar feedback.
    
    It has been found in multiple simulations by different groups that SMBHs tend to grow in two different phases \citep[e.g.,][]{dubois+2015, bonoli+2016, bower+2017, habouzit+2017, prieto+2017, daniel+2017, delayedgrowth_EAGLE+2018, Catmabacak2021}. 
    In the early universe or in low-mass galaxies, feedback by stars (in particular supernovae) regularly ejects gas from galaxy centers. 
    This results in extended periods of time during which there is little to no nearby gas for nuclear BHs to accrete. 
    In this early phase, the BH mass lags behind while the host galaxy grows its stellar mass.\footnote{It is not guaranteed that early BHs will be located at galaxy centers, e.g. if the timescale for ``sinking'' to the center is too long \citep[e.g.,][]{Ma2021_sinking}. The delayed BH growth included in our models could also, at least in part, be caused by dynamic effects such as this.}
    It is noteworthy that this result is robust to details of the BH accretion prescription used in the simulation, which are highly uncertain  \citep[see e.g.,][]{Hopkins16_concert, AA21_hyper}, as long as the accretion is tied to the gas reservoir in the immediate vicinity of the BH and the effects of stellar feedback are resolved \citep[][]{daniel+2017}. 
    The simulations find that, eventually, the gas reservoir stabilizes in galactic nuclei. 
    From that point on, nuclear BHs in star forming galaxies accrete at a much higher time-averaged rate and grow to masses comparable to those expected from local scaling relations.
    
\begin{figure*}
                \centering
                \includegraphics[width=\textwidth, trim=0cm 0.5cm 0cm 0cm, clip=true]{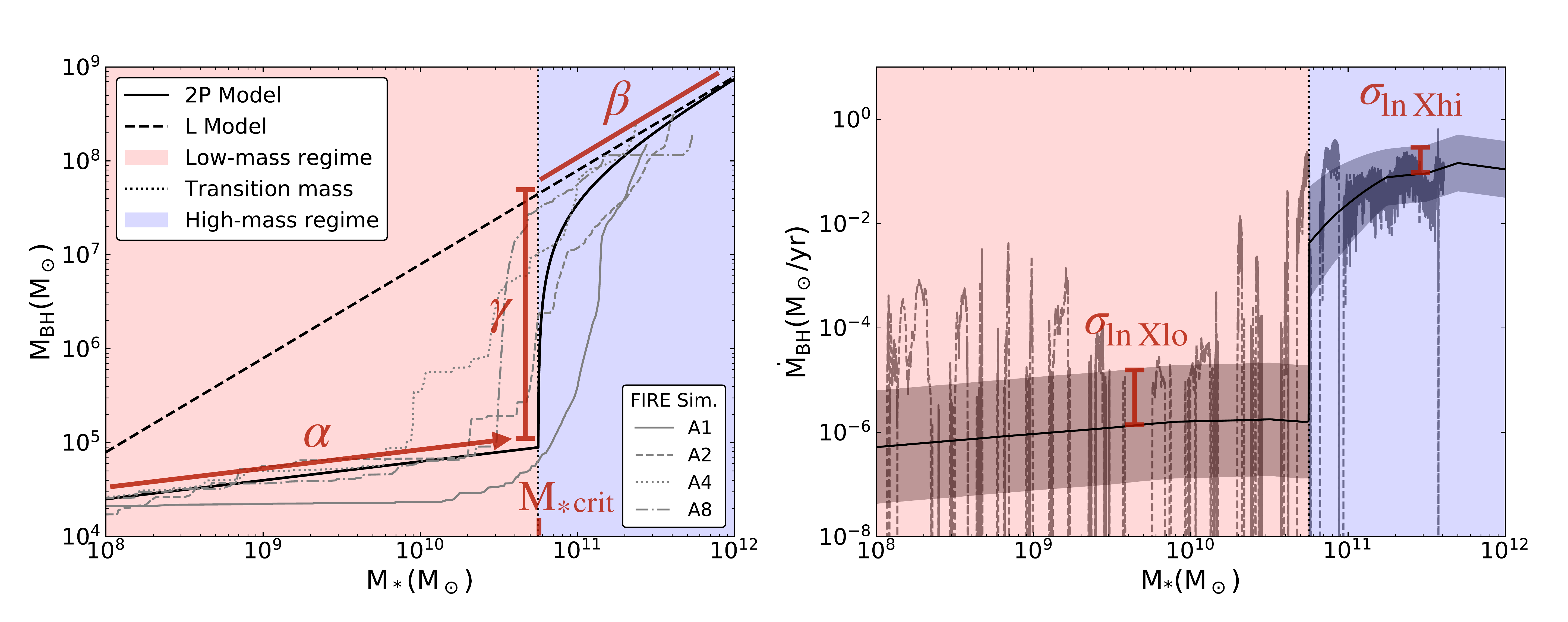}
                \caption{\textbf{Left:} The relation between BH mass and galaxy stellar mass. 
                The gray curves show the trajectories of the nuclear BHs in four different A-series simulations of massive galaxies from the FIRE project, analyzed in more depth in \citet{daniel+2017}. 
                The black curves show analytic models for the $\Mstar$-$\Mbh$ relation: a simple linear model (L; dashed) and a two-phase model (2P; solid) in which BH growth is suppressed at low masses. 
                The 2P model parameters are chosen to approximate the average trajectories of simulated BHs. 
                The red background region corresponds to the ``low-mass'' regime during which BH growth is slow, whereas the blue background region corresponds to the ``high-mass'' regime during which BH growth is more efficient. 
                The transition mass between these regimes is indicated by the vertical black dotted line.  
                \textbf{Right:} Similar but for the BH accretion rate vs. stellar mass.
                For clarity, we show the BH accretion rate for a single example simulation and only the 2P model. On each panel, the red labels indicate how each parameter of the 2P model affects the relations shown. These parameters are defined in more detail in \S \ref{s:M:ss:SMM} and in Table \ref{tab:freeparams}. 
                }\label{fig:SMBH-Relation}
            \end{figure*}

    Figure \ref{fig:SMBH-Relation} shows results from galaxy formation simulations from the FIRE (``Feedback In Realistic Environments'') project \citep[][]{hopkins+2014, hopkins+2018}\footnote{See the FIRE project web site: http://fire.northwestern.edu.} illustrating the ``two-phase" SMBH growth. 
    At the high-mass end, corresponding to later times for the galaxies tracked, the BHs end with masses roughly consistent with locally observed scaling relations (shown here in terms of $\Mbh$ vs. total galaxy stellar mass $\Mstar$). 
    However, in low-mass galaxies BHs can be under-massive relative to their host galaxies by more than an order of magnitude. 
    As Figure \ref{fig:SMBH-Relation} shows, in FIRE this produces a relation between BH and galaxy masses which has a prominent ``kink'' at a galaxy stellar mass $\Mstar \sim 5 \times 10^{10}$ M$_{\odot}$. 
    There is not yet agreement on the primary cause of the change in SMBH fueling regimes, but different possibilities have been discussed including an increase in the escape velocity in the galactic nucleus \citep[][]{dubois+2015, daniel+2017, Lapiner2021}, a change in the buoyancy of galactic winds due to formation of a hot gaseous halo \citep[][]{bower+2017}, and a change in the stability of the gaseous galactic disk, possibly owing to confinement by a hot inner circumgalactic medium (CGM; Stern et al. 2021\nocite{Stern+2020}; Gurvich et al., in prep.; Byrne et al., in prep.).
    In other simulations, the transition between BH fueling regimes also does not necessarily occur at a fixed stellar mass. For example, in EAGLE the transition is better approximated by a threshold in virial temperature of the halo \citep[][]{delayedgrowth_EAGLE+2018}.
    
    Since the relations between BH and galaxy masses are not well constrained at low masses and at high redshifts,  it is not immediately clear whether a kink at $\Mstar \sim 5 \times 10^{10}$ M$_{\odot}$ is consistent with BH mass measurements.
    In this paper, our goal is to test two-phase growth by focusing on another set of observations: the AGN luminosity function. 
    Since AGN are powered by accretion onto SMBHs, the AGN luminosity function is sensitive to the growth history of BHs \citep[e.g.,][]{Soltan_1982, SB1992, YT2002}, including the form of scaling relations as a function of redshift. 
    In practice, this is complicated by the fact that individual observations (e.g., in the optical or X-ray) typically probe only a fraction of the accretion power and by the fact that a large fraction of this accretion power can be missed entirely due to obscuration \citep[e.g.,][]{HA2018}. 
    To circumvent these difficulties, we employ
    previous studies which have modeled these effects to infer the bolometric luminosity function. We use the results from \citet{Shen2020} which updates the classic analysis of \citet{Hopkins2007} that combined a large set of AGN luminosity function measurements over the redshift interval $z=0-6$.
    These authors then obtained a bolometric luminosity function which self-consistently reproduced the observations in different bands, taking into account the luminosity dependence of intrinsic AGN spectra as well as the luminosity dependence of their obscuring columns.
    
    We want to account for the fact that the \emph{details} of the BH growth histories can differ from simulation to simulation, depending on the specific physics prescriptions used. 
    Thus, instead of comparing exact predictions from a specific set of simulations, we construct a general, parameterized semi-analytic model intended to capture a range of possible variations around the type of kinked scaling relation shown in Figure \ref{fig:SMBH-Relation}. 
    Although our model includes AGN of a wide range of luminosities, i.e. not only the more luminous AGN commonly known as quasars, we will follow the common practice of referring to the AGN luminosity function also as the quasar luminosity function, or QLF for short.
    The model scaling relation is convolved with the redshift-dependent galaxy stellar mass function, as well as a distribution of accretion rates, to derive AGN luminosity function predictions. 
    By comparing model AGN luminosity functions produced in this way with the observations, we can test whether a kinked scaling relation consistent with what is predicted by simulations is allowed or even preferred by the luminosity function data. 
    Overall, we find that a two-phase BH growth model can successfully reproduce the QLF in the redshift range $z\sim0.5-4$, where it is best constrained. 
    Although this should not be interpreted as proof due to the modeling assumptions and the limited data compared to, we furthermore find evidence that a two-phase growth model is significantly favored over a simpler model in which the relationship between BH mass and galaxy mass is linear at all masses.
    
    The plan of this paper will be as follows. \ts\ref{s:M} describes our modeling approach in more detail. 
    Model luminosity functions are fit and compared to observations in \ts\ref{s:R}. 
    We discuss the results in \ts\ref{s:D}, and summarize the main take-aways in \ts\ref{s:C}.

\section{QLF Modeling methodology}\label{s:M}
    When describing the two-phase (2P) model, we will refer to the early/low-mass and late/high-mass growth phases as the ``low-mass'' and ``high-mass'' regimes. 
    In addition to parameter variations of the 2P model, we explore a simpler model  which assumes a purely linear relation between BH mass and galaxy stellar mass (the L model). 
    By comparing best-fit 2P and L models, we can assess whether the QLF data prefer one over the other.

    \subsection{Galaxy Stellar Masses and Star Formation Rates}\label{s:M:ss:SMFSFR}
       All of our models are anchored to observations of the galaxy stellar mass function and designed such that the implied growth of the BH population is consistent with the observed growth of the host galaxy population. 
       We utilize Universe Machine \citep[UM,][]{behroozi+2019} to model galaxy properties as a function of redshift. 
        In particular, we use UM to model the redshift-dependent galaxy stellar mass function (SMF) and the mean specific star formation rates (sSFRs) of galaxies. 
        In UM, these properties are self-consistently constrained based on a wide range of observations, including different luminosity and correlation functions. 
        UM provides the SMF and sSFR data for a large number of stellar mass and redshift bins. In our modeling code, we interpolate smoothly between the values provided.
    
    \subsection{The Two-Phase and Linear Models}\label{s:M:ss:SMM}
    
    In this section, we describe in more detail how we implement the two-phase model. 
    The linear model is simply a special case of the 2P model in which there is no transition between distinct regimes.
    
    \subsubsection*{Black Hole Masses}\label{s:M:sss:SMBH}
    
    In the 2P model, BHs are assumed to follow, on average, a relation between BH mass and total galaxy stellar mass that has two distinct phases, corresponding to the low-mass and high-mass regimes: 
            \begin{equation}\label{eqn:SMBH-piece}
                \Mbh = 
                \begin{cases}
                \Mbhpre & \Mstar < \MstarCRIT \\ \\
                \Mbhpost & \Mstar > \MstarCRIT,
                \end{cases}
            \end{equation}
    where the two regimes are separated by a critical stellar mass $\MstarCRIT$ (the `transition mass'). 
    The BH mass scalings in the two regimes are parameterized by three dimensionless parameters: 
    the power-law slope in the low-mass regime ($\alpha$), the asymptotic $M_{\rm BH}/M_{*}$ ratio at high mass ($\beta$), and the factor by which the BH is undermassive at $\MstarCRIT$ relative to the linear relation ($\gamma$). 
    
    In the low-mass regime, when BHs are undermassive, we use the following parameterization to describe the scaling relation:
    \begin{equation}\label{eqn:pre-disk-kink}
        \Mbhpre = \MbhCRIT \left( \frac{\Mstar}{\MstarCRIT} \right) ^{\preslope},
    \end{equation}
    where $\MbhCRIT$ sets the normalization in the low-mass regime. 
    For the high-mass regime, we set the requirement that the relation approaches linearity, 
    \begin{equation}
    \Mbhpost(\Mstar >> \MstarCRIT) \to \beta \Mstar,
    \end{equation}
    but we do not enforce a strictly linear relation between $M_{\rm BH}$ and $M_{*}$. 
    The linear slope at high masses approximates the scaling relations often found in observations \citep[e.g.,][]{marconi+2003, haring+2004, kormendy&ho2013}. 
    Rather, we use continuity considerations described in more detail below to connect the low- and high-mass regimes. 
    We then introduce the dimensionless parameter $\gamma$ to set the normalization of the low-mass scaling relation,
    \begin{equation}
    \label{eqn:gamma_def}
        \gamma = \frac{\beta \MstarCRIT}{\MbhCRIT}.
    \end{equation}
    
    To obtain a functional form for $\Mbhpost$ that connects to the low-mass regime, we make the ansatz that as soon as galaxies enter the high-mass regime, the BH mass and stellar mass of the galaxy start growing in proportion to each other:
    \begin{equation}\label{eqn:proportional-growth}
        \MdotBH = \postnorm \MdotStar,
    \end{equation}
    implying that in the high-mass regime, 
    \begin{equation}\label{eqn:post-disk-kink}
        \Mbhpost = \MbhCRIT + \postnorm \left( \Mstar - \MstarCRIT \right).
    \end{equation}  
    Figure \ref{fig:SMBH-Relation} shows an example of the $\Mbh$-$\Mstar$ relation from the 2P model, overplotted on FIRE simulation data.         
    
            We note that in observations $\Mbh$ has often been found to correlate more tightly with the bulge mass rather than the total stellar mass of the galaxy. 
            We use total stellar mass in our model because the simulations predict the 2P growth behavior in the $\Mbh$-$\Mstar$ relation (see Fig. \ref{fig:SMBH-Relation} and Byrne et al., in prep.) and not only when $\Mbh$ is plotted as a function of a proxy for bulge mass \citep[][]{daniel+2017}. 
            We did however explore model variations in which we use bulge mass and found that our main conclusions regarding the two-phase model are not changed. 
            This makes sense because, at least at the low redshift where detailed bulge/disk decompositions are available, the bulge mass on average dominates the total stellar mass above $\sim \MstarCRIT$ \citep[e.g.,][]{benson+2007}. 
            Thus, the effects of distinguishing between bulge mass and total stellar mass are partially degenerate with the parameters that describe the break in the $\Mstar-\Mbh$ relation.

        \begin{table}
            \centering
            \renewcommand{\arraystretch}{1.2}
            \begin{tabular}{m{2.65cm}>{\raggedright\arraybackslash}m{5.25cm}}
            Parameter & Description \\ \hline \hline
            \begin{tabular}{c}
                $\MstarCRIT$ \\ (transition mass)
                \end{tabular} & The stellar mass at which the low-mass regime ends and the high-mass regime begins. 
                \\ \hline
            \begin{tabular}{c} 
                $\preslope$ \\ (low-mass slope) 
                \end{tabular} &  The power-law slope of the $\Mstar$-$\Mbh$ relation in the low-mass regime.
                \\ \hline
            \begin{tabular}{c}
                $\postnorm$ \\ (high-mass \\ normalization)
                \end{tabular} & The asymptotic $\Mbh/\Mstar$ ratio in the high-mass regime.
                \\ \hline
            \begin{tabular}{c}
                $\gamma$ \\ (break factor)
                \end{tabular} & The factor by which the BH is undermassive at $\MstarCRIT$ relative to the high-mass power law. 
                \\ \hline
            \begin{tabular}{c}
                $\siglnXpre$ \\ (low-mass $\siglnX$)
                \end{tabular}& The log-normal standard deviation of the normalized BH accretion rate distribution in the low-mass regime. 
                \\  \hline
            \begin{tabular}{c}
                $\siglnXpost$ \\ (high-mass $\siglnX$)
                \end{tabular} & Same as above but for the high-mass regime.
                \end{tabular}
                \caption{Overview of the two-phase (2P) model parameters.}\label{tab:freeparams}
            \end{table}

        \begin{figure*}
            \centering
            \includegraphics[width = \linewidth, trim=0.0cm 0.0cm 0.0cm 0.0cm, clip=true]{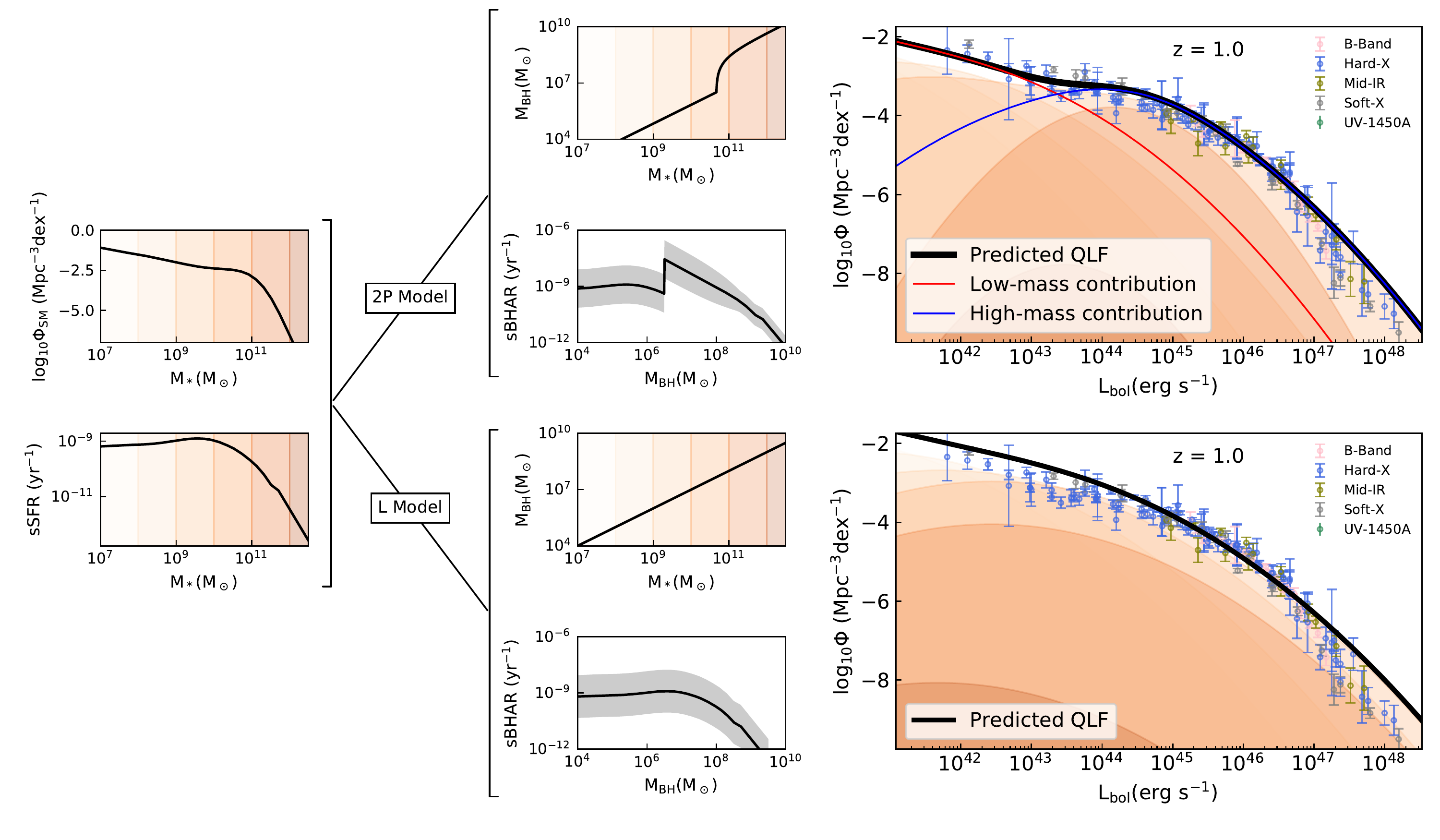}
            \caption{Illustration of different ingredients used in producing model QLFs, at $z=1$. 
            \textbf{Left column:} The models are anchored to the observed galaxy stellar mass function and the self-consistently derived mean specific star formation rates of galaxies, which we model using Universe Machine. 
            The top panel shows the SMF (units of $\tn{Mpc}^{-3} \log_{10}[\Mstar]^{-1}$) and the bottom panel shows the mean sSFR vs. stellar mass.
            \textbf{Middle column:} The top branch shows the $\Mbh-\Mstar$ relation for a two-phase model with a break at $\MstarCRIT$ and the corresponding mean specific BH accretion rate vs. BH mass. 
            The bottom branch shows the same quantities but for a purely linear model (no break in $\Mbh-\Mstar$). 
            \textbf{Right column:} The resulting bolometric QLFs (in units of $\tn{Mpc}^{-3} \log_{10} \left[\Lbol\right]^{-1}$) for the two-phase and linear models. 
            The solid black curves show the best-fit model QLFs at $z=1$ when the model parameters are assumed to be independent of redshift (fixed) and the models are simultaneously fit to observations at $z=0.5,~1,~2,~3,$ and 4 (see \S \ref{sec:2P_model_fits}). 
            The solid red and blue curves in the top panel correspond to the low-mass and high-mass contributions to the two-phase QLF, respectively. The orange shaded regions correspond to different stellar mass bins which can be seen in the left column of this figure.
            The data points are from the \citet{Shen2020} bolometric QLF and correspond to inferences for observations in different bands.}
            \label{fig:QLF_rep}
        \end{figure*}

        \subsubsection*{Black Hole Accretion Rates}\label{s:M:sss:X}
  
            Next we describe how we model the distribution of BH accretion rates. 
            At any given redshift, the distribution of BH masses is obtained by convolving the galaxy stellar mass function with the $\Mbh-\Mstar$ model from the previous section. 
            Since the galaxy stellar mass function evolves with redshift, this implies a specific redshift evolution for the BH mass function. 
            To constrain the distribution of accretion rates, we enforce the requirement that the mean BH accretion rate for any given stellar mass and redshift is consistent with the time evolution of the BH mass function.
            
            The BH accretion rate can be related to the stellar mass growth rate via the following identity: 
            \begin{equation}\label{eqn:MdotBH}
                \frac{d \Mbh}{dt} = \frac{\Mbh}{\Mstar} \frac{d\ln \Mbh}{d\ln \Mstar} \frac{d \Mstar}{dt}.
            \end{equation}
            Averaging both sides of this equation (treating $\Mstar$ as the independent variable and the slope of the scaling relation as a constant at fixed $\Mstar$) and combining with equation (\ref{eqn:SMBH-piece}) we obtain for the $\MdotBH-\MdotStar$ relation:
            \begin{equation}\label{eqn:MdotBH-piece}
                \expect{\MdotBH} = 
                \begin{cases}
                \preslope \Mbhpre \frac{\expect{\MdotStar}}{\Mstar} & \Mstar < \MstarCRIT \\ \\
                \postnorm \expect{\MdotStar} & \Mstar > \MstarCRIT
                \end{cases}.
            \end{equation}
            This form is convenient because the terms involving the galaxy stellar mass and its rate of growth can be self-consistently modeled using results from UM. 
            In doing so, we assume that the factor $\expect{\MdotStar}/\Mstar$ above equals the mean sSFR as a function of stellar mass and redshift which we calculate from UM.
            \footnote{This identification neglects a $\sim 20$\% difference between the SFR and the net stellar mass growth rate $\MdotStar$ owing to stellar mass loss \citep[e.g.,][]{Leitner2011}. 
            We neglect this difference because it is degenerate with the normalization of the scaling relation, which is a free parameter. 
            We also neglect the fact that the stellar mass of a galaxy can grow via mergers. 
            This is a fair approximation for our purposes because this primarily affects the most massive galaxies \citep[e.g.,][]{AA17_cycle} and we do not expect this to significantly change the possible signatures of a change in BH accretion properties at intermediate stellar masses.}
            We note that, using the relations in the previous sections, the other terms can be fully parameterized by $\MstarCRIT$ and the dimensionless parameters $\alpha$, $\beta$, and $\gamma$.

            To account for the strong variability in BH accretion rates, we define a distribution of accretion rates around the mean. 
            We use a log-normal distribution for simplicity and to capture a wide range of accretion rates. 
            To minimize the number of free parameters, we make the ansatz that the distribution of BH accretion rates can be parameterized by the distribution of dimensionless fluctuations
            \begin{equation}
                \tn{X} \equiv \frac{\MdotBH}{\expect{\MdotBH}}.
            \end{equation}
            The X distribution is assumed to be independent of redshift and to depend only on whether the galaxy is in the low-mass regime or the high-mass regime.
            The log-normal distribution is then defined as:
            \begin{equation}\label{eqn:X-PDF}
                p \left[ \ln \tn{X} \right] =  \frac{1}{\sqrt{2 \pi} \siglnX} \  \tn{exp} \left( \frac{- \left( \ln \tn{X} - \mulnX \right)^2}{2 \siglnX^2} \right).
            \end{equation}
            Using the requirement that by definition $\expect{\tn{X}}\equiv 1$, implying $\mulnX = - 0.5 \siglnX^2$, the distribution is fully characterized by the single parameter $\siglnX$. 

            The low-mass and high-mass distributions parameters are labeled $\siglnXpre$ and $\siglnXpost$, respectively. 
            Since accretion is more sporadic in the low-mass regime, we expect that $\siglnXpre \geq \siglnXpost$ and impose this requirement when fitting the model to observational data (see \S \ref{s:M:ss:BF}). 
            Furthermore, to avoid a discontinuity in $\siglnX$ at $\MstarCRIT$, the log-normal dispersion is implemented such that it changes continuously from $\siglnXpre$ at $\Mstar\leq\MstarCRIT$ to $\siglnXpost$ at a stellar mass 0.5 dex higher than  $\MstarCRIT$. 
            The interpolation is done linearly in $\siglnX$ vs. $\log{\Mstar}$ space.
            
            To further characterize BH growth in the models, we define the mean specific BH accretion rate ${\rm sBHAR} \equiv \expect{\MdotBH}/\Mbh$, which is a function of redshift and BH mass. 
            Note that this quantity is related to the Eddington ratio, but it is expressed in different units and averaged over the BH population, for a given $\Mbh$.
            
\begin{figure*}
            \centering
            \includegraphics[width=\linewidth, trim=0.425cm 0.5cm 0.37cm .2cm, clip=true]{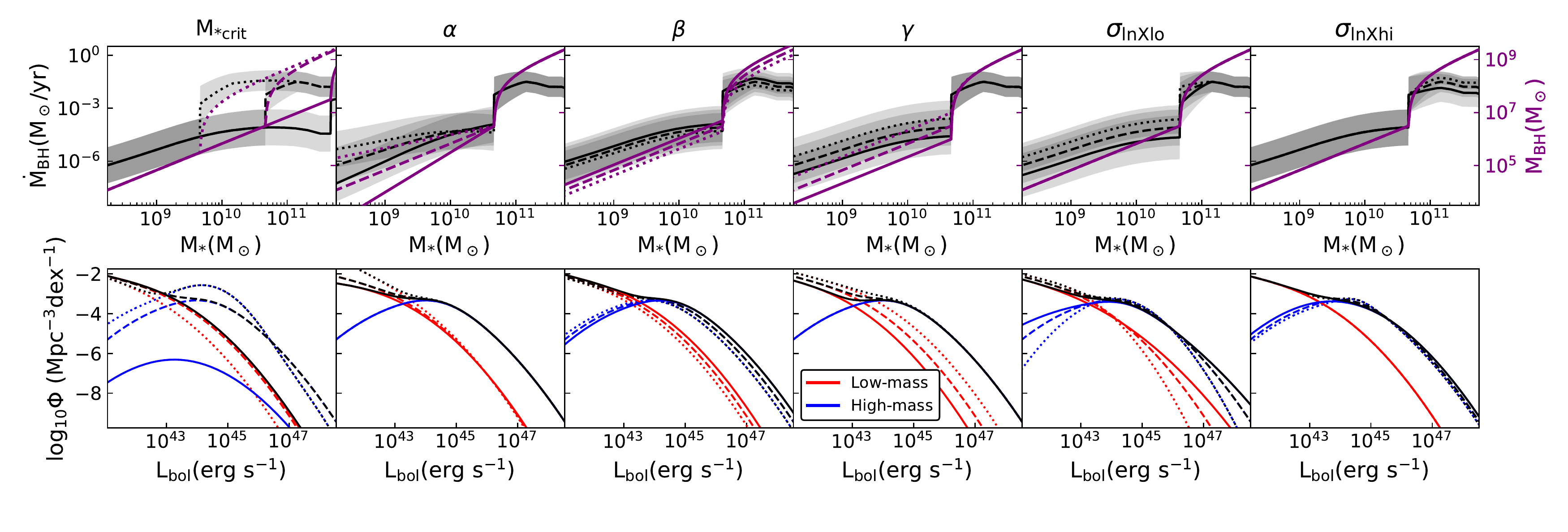} 
            \caption{
            Illustration of how varying each of the six free parameters of the two-phase model affects the results, for the representative redshift $z=1$ (see Table \ref{tab:freeparams} for parameter definitions). 
            \textbf{Top row:} Effects on the mean $\Mstar-\MdotBH$ (black) and $\Mstar-\Mbh$ (purple) relationships. 
            The gray region corresponds to the $\pm \siglnX$ range that characterizes the log-normal distribution of accretion rates.
            \textbf{Bottom row:} Effects on the resulting QLF. 
            The black curves correspond to the total QLF, while the red and blue curves show contributions from the low-mass and high-mass regimes, respectively. 
            The values of the free parameters that are varied are, from left to right: $\MstarCRIT=(10^{9.67}, 10^{10.67}, 10^{11.67} )~\solmass$, $\alpha=(0.5, 1.0, 1.5)$, $\beta=(10^{-2.53}, 10^{-2.33},\ 10^{-2.13})$, $\gamma=(10^{1.34}, 10^{1.84}, 10^{2.34})$, $\siglnXpre=(1.85, 2.35, 2.85)$, and $\siglnXpost=(0.83, 1.33, 1.83)$. 
            The results for the different parameter values are denoted by different line styles respectively as (dotted, dashed, solid). 
            }
            \label{fig:freeparams}
        \end{figure*}
        
    \subsubsection*{The Linear Limit}\label{s:M:ss:SPL}            
        To assess whether two-phase SMBH growth is preferred by the QLF data, we also explore a model in which the relation between BH mass and stellar mass is purely linear and with no transition in the normalized accretion rate distribution, i.e. with constant $\siglnX$. 
        This linear model is fully specified by the $\beta=M_{\rm BH}/\Mstar$ ratio and a single log-normal accretion rate dispersion $\siglnX$. 
        We note that, as for the 2P model, our L model is self-consistently anchored to the redshift-dependent SMF and sSFR data from UM. 
        This is significant because some previous QLF models also assumed a log-normal accretion rate distribution at any given mass and redshift, but did not include the requirement that the distribution of accretion rates must self-consistently connect the implied BH mass functions at different redshifts (e.g., Conroy \& White 2013; though see Veale et al. 2014)\nocite{conroy+white2013, veale+2014}.
        
    \subsection{Producing the QLF}\label{s:M:ss:C}
    The steps described so far yield a distribution of BH accretion rates for any given stellar mass and redshift. 
    We convert this straightforwardly to a bolometric luminosity distribution by assuming a constant radiative efficiency:
    \begin{equation}\label{eqn:MdotBH-Lbol}
            \Lbol = \epsilon \MdotBH \tn{c}^2.
        \end{equation}
    For simplicity we set $\epsilon=0.1$ to represent radiatively efficient accretion disks \citep[e.g.,][]{Abramowicz+2013}. 

    The observed QLF includes accretion in galaxies of all stellar masses, so to obtain the luminosity function at any redshift, we integrate over the galaxy SMF:
    \begin{equation}\label{eqn:QLF}
        \frac{d\tn{N}}{d\ln\Lbol} = \int p \left[ \ln \Lbol | \ln \Mstar \right] \frac{d\tn{N}}{d\ln\Mstar} d \ln \Mstar,
    \end{equation}
    where 
    \begin{equation}\label{eqn:QLF-PDF}
        \resizebox{.88\hsize}{!}{$p \left[ \ln \Lbol | \ln \Mstar \right] = \frac{1}{\sqrt{2  \pi \siglnMdot^2}} \tn{exp} \left( \frac{- \left( \ln \MdotBH - \mulnMdot \right)^2}{2 \siglnMdot^2} \right)$,}
    \end{equation}
    \begin{equation*}
        \mulnMdot = \mulnX + \ln \expect{\MdotBH},
    \end{equation*}
    \noindent and
    \begin{equation*}
        \siglnMdot = \siglnX.
    \end{equation*}
    
        We do not explicitly model radiatively inefficient accretion, based on the assumption that doing so would only significantly affect the predicted QLF at luminosities too low to affect our conclusions regarding the effects of a change in BH fueling at $\sim \MstarCRIT$.
        We also do not explicitly model scatter in the $\Mbh-\Mstar$ relation. 
        Instead, we assume that the net effect of such scatter can be modeled implicitly as a contribution to the scatter in the accretion rate distribution.
        
        Figure \ref{fig:QLF_rep} illustrates our process to produce model QLFs, using $z=1$ data. 
        The leftmost column displays the SMF and sSFR data.
        The middle column shows example $\Mstar$-$\Mbh$ and sBHAR relations for the 2P model (top) and the linear model (bottom). 
        The rightmost column shows the implied QLF for each model.   
        The panel for the 2P model additionally shows the QLF contributions from the low-mass and high-mass regimes. 
        For this figure, the model parameters are best fits to the \cite{Shen2020} QLF data when parameters are assumed to be independent of redshift (fixed) and the models are simultaneously fit to observations at $z=0.5,~1,~2,~3,$ and 4 (we describe the different fits we explore in the next section). 
        
        Figure \ref{fig:freeparams} shows the effects the different free parameters have on the predicted QLF, as well as the $\Mstar$-$\Mbh$ relation and the $\Mstar$-$\MdotBH$ relations. 
        The effects on the predicted QLF are mostly as may be expected. 
        For example, increasing $\beta$ increases BH masses and BH accretion rates linearly in the high-mass regime, so this increases the QLF at the high end. 
        Increasing the break factor $\gamma$ decreases the same quantities in the low-mass regime, so this tends to decrease the QLF at the low end, and can help with imprinting a knee shape (in phenomenological QLF fits, the ``knee'' is where there is a break in power-law slope between the low-luminosity and high-luminosity regimes). 
        Interestingly, we see that the scatter in the accretion rate distribution (via the $\siglnX$ parameters) is important in determining the number of high-luminosity quasars (as has also been found in previous studies, e.g. Veale et al. 2014)\nocite{veale+2014}. 
        We note that varying the $\siglnX$ parameters changes the mean accretion rates because the accretion rate distribution is normal in the logarithm, so the linear mean shifts with $\siglnX$.

\begin{figure*}
        \centering
        \includegraphics[width=\linewidth, trim=0cm 0cm 0cm 0.0cm, clip=true]{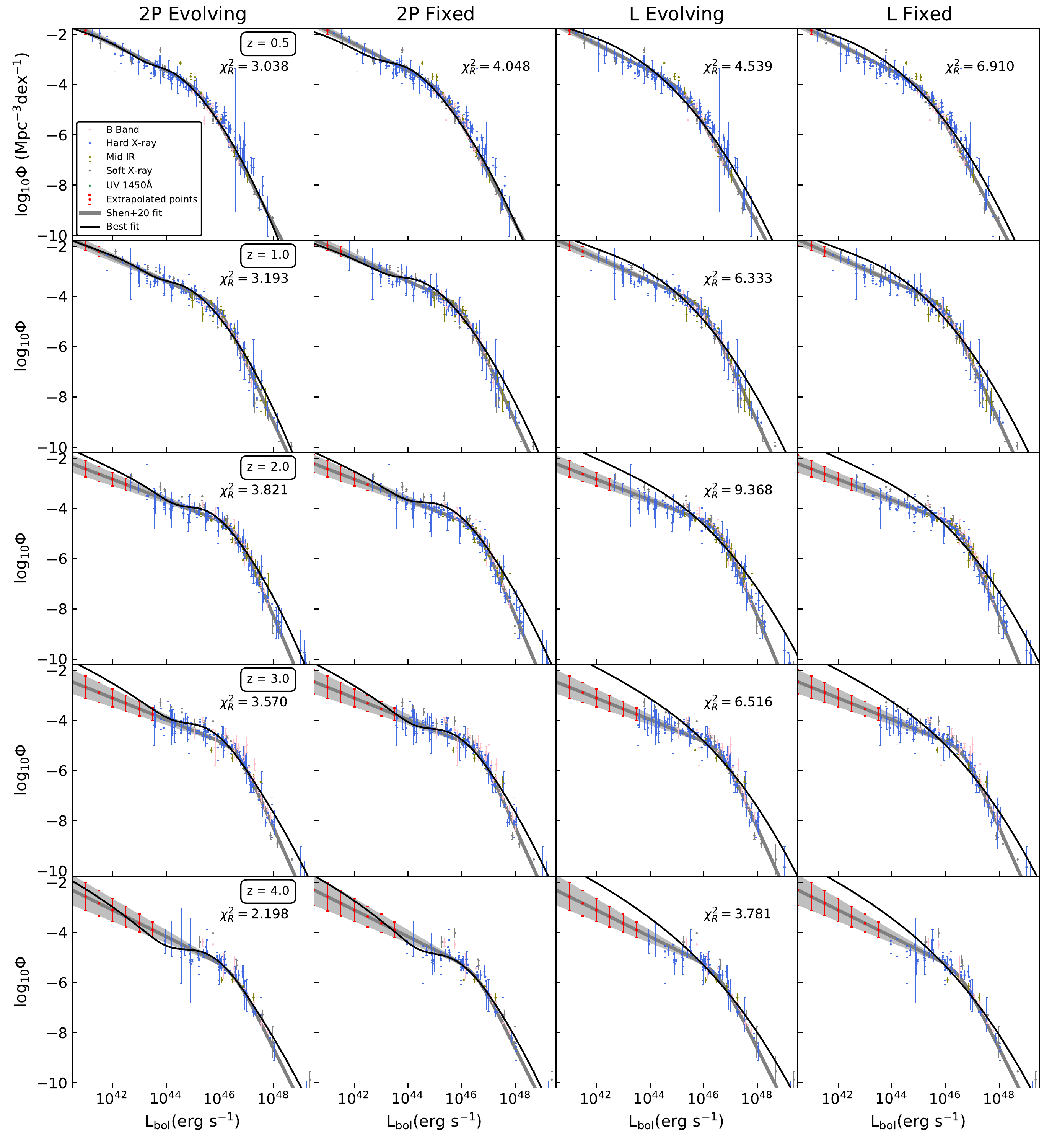}
        \caption{Best-fit QLFs for the two-phase (2P) and linear (L) models, in solid black. \textbf{1st column:} 2P model with parameters allowed to vary with redshift (evolving). \textbf{2nd column:} 2P model but with fixed parameters fit simultaneously to all redshifts shown. \textbf{3rd column:} L model with evolving parameters. \textbf{4th column:} L model with fixed parameters.
Each row displays results for the redshift indicated in the 1st column.
The models are fit to bolometric QLF values inferred by \citet{Shen2020} based on observations in different bands, plus extrapolated points at the low-luminosity end (in red) to penalize models that predict a change in slope where observations are not available. The best-fit double power law model from \citet{Shen2020} is shown in gray for reference (their ``Global A'' fit).
The $\chi^2_R$ quantity shown in panels is the reduced $\chi^2$ value for the plotted fit.
For the fixed fit, there is a single $\chi^2_R$ corresponding to the global fit.
As discussed in the text, the 2P fits shown here assume a fixed $\alpha=1$ value because it is found that freeing this parameter does not significantly improve the quality of the fits.}
        \label{fig:QLF_bestfits}
    \end{figure*}

\section{Results}\label{s:R}
    \subsection{Fitting to Observational Data}\label{s:M:ss:BF}

        We conduct fits of both our 2P and L model QLFs to observational data on the bolometric QLF compiled in \citet{Shen2020}. Their study compiles multi-wavelength observational data, including updates from the last decade, using a quasar SED model and bolometric/extinction corrections to update constraints on the observed bolometric QLF from the earlier \citet{Hopkins2007} study. \citet{Shen2020} provides constraints on the QLF from $z = 0$ to $z = 7$, but in our study we focus our fits on $z=0.5,~1,~2,~3$ and $4$ since the observational data is most complete within this range.
        
        To find best-fit parameters, we employ a least-squares method over a finite parameter space. 
        The goodness-of-fit is evaluated at each point in parameter space using the reduced  $\chi^2$ statistic computed in the standard way, $\chi^2_R = \chi^2/(N-p)$, where $N$ is the number of data points in the fit and $p$ is the number of free parameters. 
        We evaluate $\chi^2_R$ over a multidimensional Cartesian grid and find the best fit by minimizing $\chi^2_R$. 
        
        We carried out several versions of the fit. 
        For each of the 2P and L models, we tried both fitting the model parameters to each redshift independently (the ``evolving'' fits) and requiring a fixed set of parameter values to simultaneously fit the data at all redshifts (the ``fixed'' fits).
        For the evolving fits, $\chi^2_R$ is evaluated for each redshift, but for the fixed fits, a single $\chi^2_R$ is evaluated which includes data at all redshifts.
        
        For the 2P model, we conducted one fit with all six model parameters (summarized in Table \ref{tab:freeparams}) free, and another with only five free parameters in which $\alpha$ was fixed to a value of $1$.
        The motivation for the latter fit is that $\alpha$ parameterizes the low-mass slope of the $M_{\rm BH}-M_{*}$ relation and may not be well constrained because of luminosity limits on the observations. 
        We found that the reduced $\chi^2$ values were nearly identical for the fits with $\alpha$ free in the range $0 \leq \preslope \leq 2$ (which resulted in a best fit value of $\alpha = 0.9$) versus fixed at $\alpha = 1$. 
        For this reason, and to minimize degeneracies associated with multiple parameters, we focus on 2P results for $\alpha=1$ fixed for the rest of the paper. 
        Thus, the 2P fits that follow have five free parameters while our linear fits have two free parameters.
    
    	Our approach is to let the fits explore wide ranges of parameter values (i.e., to avoid prescribing constraining priors), so that we can determine the parameter values favored by the QLF data alone. 
    	The range considered for $\beta$ was broadly motivated by BH mass measurements at the high end, but nevertheless set wide enough to allow the QLF data to determine a favored value. 
    	We confirmed that the likelihood for each free parameter peaks well within the explored range. 
    	The parameter ranges used for our final fits are (in the notation for the 2P model): $8 \leq \log_{10}{\MstarCRIT} \leq 12$, $-3.1 \leq \log_{10}{\postnorm} \leq -1.8$, $0 \leq \log_{10}{\gamma} \leq 4$, $0 \leq \siglnXpre \leq 5$, and $0 \leq \siglnXpost \leq 5$. 
        When fitting the 2P model, we implement a physical prior requiring $\siglnXpost \leq \siglnXpre$. 
        This is because in hydrodynamical simulations, the accretion rate variability is predicted to be higher in the low-mass regime.  

        In addition to fitting the observation data points from \citet{Shen2020}, we include in the fits additional ``extrapolated points'' at the low-luminosity end at each redshift. 
        The extrapolated data points are intended to penalize models that imply a low-luminosity QLF shape that diverges strongly from the usual power-law form, which is found where low-luminosity data are available (e.g., down to $L_{\rm bol}<10^{42}$ erg s$^{-1}$ at $z=0.5$). 
        The extrapolated points are most important at high redshift, where the observations do not directly constrain the low-luminosity end (e.g., only down to $L_{\rm bol}\sim10^{44}$ erg s$^{-1}$ at $z=4$). 
        For each redshift, the extrapolated points are introduced starting at the low luminosity limit of the observational data (for the given $z$), and continue every 0.5 dex down to $\Lbol = 10^{41}$ erg s$^{-1}$. 
        We assume these points lie on the \citet{Shen2020} ``Global A'' best fit and the uncertainty is calculated based on the uncertainty of the Global A fit's parameters.
        
    \subsection{QLF Fit Results}
    \label{sec:2P_model_fits}
    
        Figure \ref{fig:QLF_bestfits} shows the best-fit QLFs for both the 2P and L models, and for the evolving vs. fixed fits. 
        The $\chi^2_R$ values for the fits are indicated on the figure panels (for the evolving fit, there is a $\chi^2_R$ for each redshift, but for the fixed fits, there is a single value for the global fit to all redshifts simultaneously, indicated in the top row). 
        
        We must first acknowledge that none of the fits are ideal from a statistical point of view because the $\chi^2_R$ values are all significantly above unity. 
        There are a couple likely reasons for this. 
        One is that a close examination of the observational data points suggests that not all the data points are consistent with each other. 
        For example, there are mid-IR and soft X-ray data points with small error bars that appear systematically above other observations and above the \citet{Shen2020} best fit (e.g., around the knee of the QLF at $z=0.5$). 
        This suggests that there are some systematic effects not accounted for in the error bars, and that even a perfect model would produce a fit with a $\chi^2_R$ exceeding unity. 
        In fact, our evolving 2P fit appears to describe very well the $z=0.5$ observed QLF over the entire luminosity range plotted, even though it has a $\chi^2_R=3.038$. 
        We also note that our models are relatively simple and that there is no a priori guarantee that they can capture all the complexities of the observed data. 
        In particular, our 2P model includes the minimum number of parameters necessary to describe a two-phase scenario in which the parameter values are not fixed.
        Both the 2P and L models furthermore assume a simple log-normal distribution of normalized accretion rates (eq. \ref{eqn:X-PDF}), which may not be a fully accurate characterization of AGN activity. 
        Nevertheless, the fits are useful to assess the degree to which a two-phase model is favored over a linear model. 
        
        Figure \ref{fig:QLF_bestfits} shows that 2P fits are systematically better than the L fits. 
        This is the case for the fixed fits, as well as for the evolving fits for each redshift. 
        For the fixed fits, the difference in reduced $\chi^2$ is $\Delta \chi^2_R = 6.910-4.048=2.862$ in favor of the 2P model. 
        As mentioned above, the 2P evolving fit appears to describe the $z=0.5$ observations very well, and this is the redshift for which the observations cover the largest luminosity range. 
        At higher redshifts $z\geq2$, the 2P fits imply a ``bumpy'' QLF shape and a low-luminosity end with a steeper slope than the empirical double power-law fit from \citet{Shen2020}. 
        We note, however, that these effects are seen where low-luminosity data become sparse or non-existent (other than through the extrapolated points), so this may be largely due to the poor constraints on the fits. 
        We therefore do not consider the ``bumpy'' low-luminosity QLF shape favored by some 2P fits in higher-redshift bins to be a robust prediction of two-phase growth. 
        On the other hand, the L fits fail in a generic way to describe the observed QLF because they are unable to reproduce a clear knee. 
        This is especially evident at intermediate redshifts $z=2-3$, where the observations show a distinct knee in the QLF which the best-fit L models entirely fail to capture. 
        
\begin{figure}
            \centering
            \includegraphics[width=\linewidth, trim=0cm 0cm 0cm 0cm, clip=true]{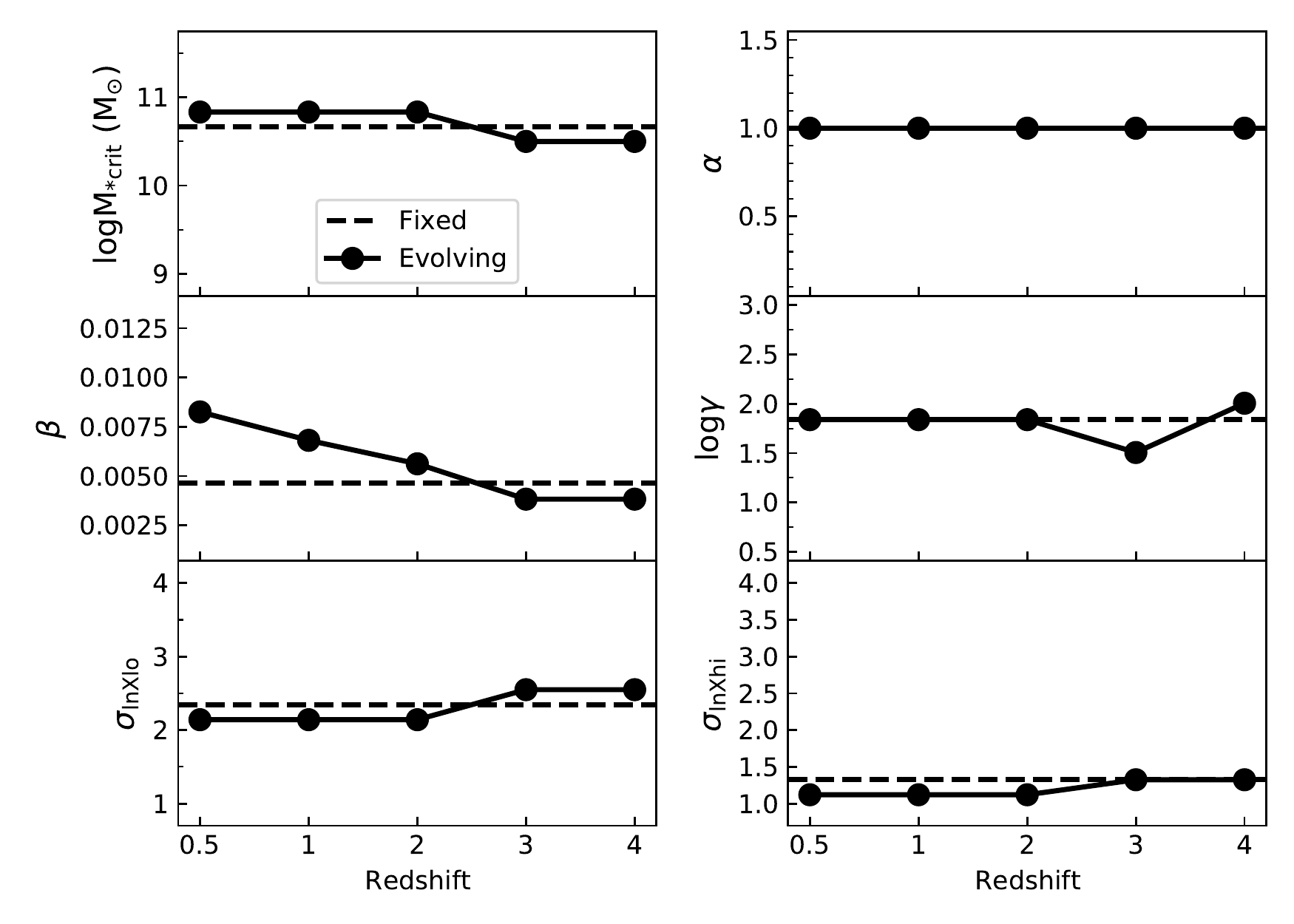}
            \caption{Comparison of the best-fit parameters for the two-phase model when the parameter values are allowed to evolve with redshift (black dots) vs. are fixed (dashed lines). Overall, the best-fit parameter are relatively stable with redshift even when allowed to vary, though some evolution in $\beta$ (the high-mass $\Mbh$-$\Mstar$ normalization) is suggested by the fits.
As in Fig. \ref{fig:QLF_bestfits}, the fits shown here assume a constant $\alpha=1$.}
            \label{fig:bestfit_params}
        \end{figure}

        Figure \ref{fig:bestfit_params} summarizes the best-fit parameters for the 2P model, and shows how the evolving parameters compare to the fixed parameters. 
        This provides a useful check on the 2P model, because physically we expect that if the model is a good description of the AGN population, the model parameters should evolve smoothly with redshift. 
        This is supported by the results in the figure, which show that the best-fit evolving parameters are either stable or only modestly evolving with redshift. 
        This result also explains why the fixed fits are only slightly worse than the evolving fits. 
        Interestingly, the evolving fits suggest some evolution in $\beta$ (the high-mass $\Mbh$-$\Mstar$ normalization), such that $\Mbh/\Mstar$ increases by a factor $\sim 2$ from $z\sim3-4$ to $z=0.5$. 
        An increase in the high-mass $\Mbh/\Mstar$ ratio by a comparable amount over this redshift interval is also suggested by the recent Trinity empirical model \citep[][]{Zhang2021_Trinity}.
        
\begin{figure*}
            \centering
            \includegraphics[width=\linewidth, trim=0cm 0cm 0cm 0cm, clip=true]{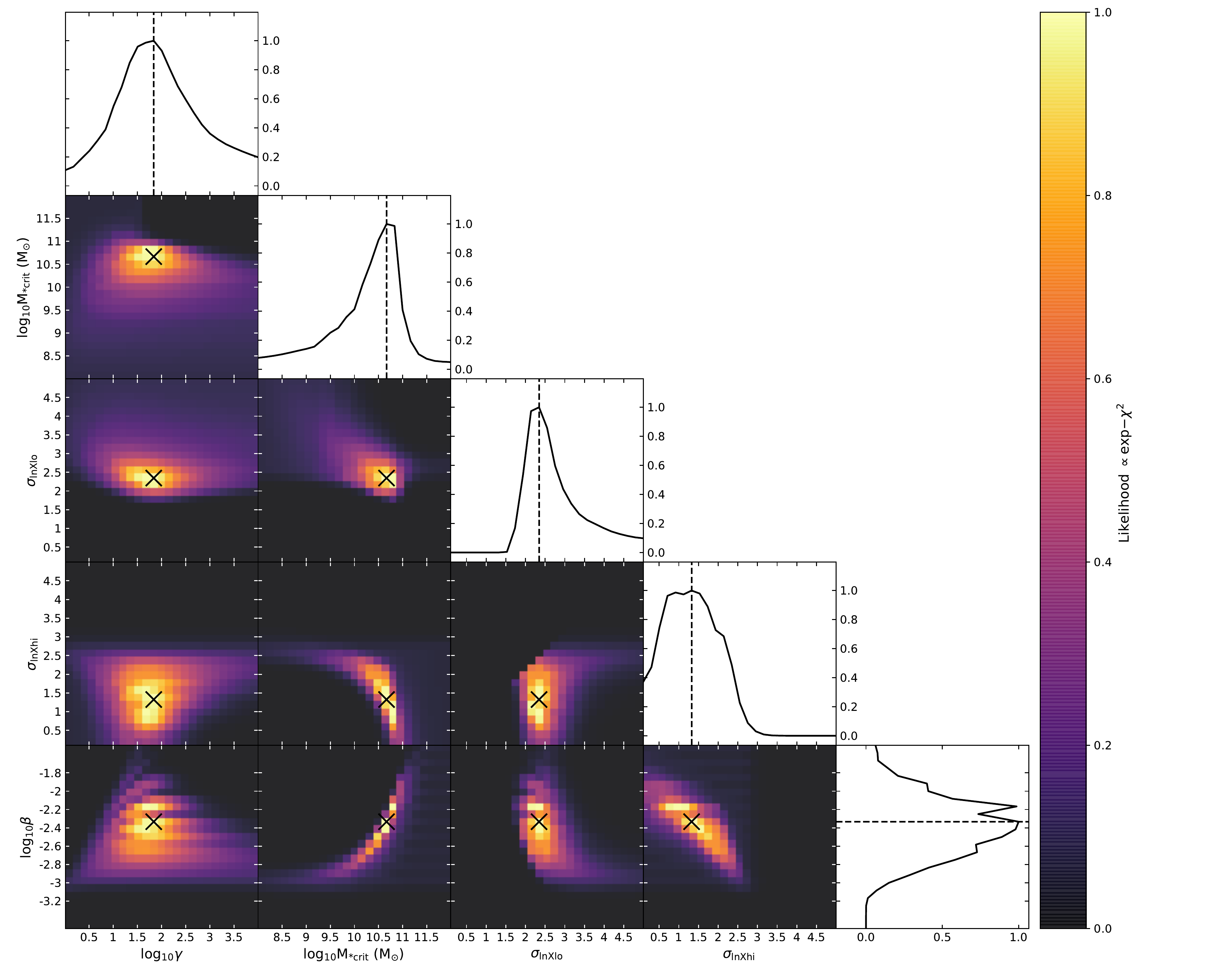}
            \caption{Corner plot displaying the bi-variate normalized likelihood contours for the two-phase model.
This is for the fit with parameters fixed with redshift, and with fixed $\alpha=1$ as in Figs. \ref{fig:QLF_bestfits} and \ref{fig:bestfit_params}.
The 1D panels show profile likelihoods (which indicate the maximum value of the likelihood function for the parameter, with all other parameters free) for the five free parameters.
The fit to the observational data favors a transition mass $\MstarCRIT \approx 10^{10.7} \solmass\approx 5\times10^{10} \solmass$ and a break factor $\gamma \approx 10^{1.8}\approx60$ broadly consistent with physical expectations for a two-phase model from simulations.}
            \label{fig:chi2}
        \end{figure*}

        For reference, the best-fit values for the fixed 2P model are $\log_{10} \MstarCRIT = 10.7 ^{+0.3}_{-0.5} \solmass$, $\log_{10}\postnorm = -2.3^{+0.3}_{-0.4}$, $\log_{10} \MbhNORM = 1.8^{+0.6}_{-0.8}$, $\siglnXpre = 2.4^{+0.4}_{-0.4}$, and $\siglnXpost = 1.3 ^{+0.9}_{-1}$. The best-fit values for the fixed L model are $\log_{10}\postnorm = -3.00^{+0.08}_{-0.08}$ and $\siglnX = 2.6^{+0.2}_{-0.2}$. The errors are $1\sigma$ and determined by where the 1D profile likelihood for each parameter drops to $e^{-1/2}$ of its peak value (see the 1D likelihood panels in Fig. \ref{fig:chi2}).
    
\section{Discussion}\label{s:D}

    In the previous section, we showed that a 2P model fits the QLF data better than an L model. 
    In this section, we assess in more detail the evidence that the QLF data supports two-phase SMBH growth. 
    
    One potential concern is that the 2P model might fit the QLF observations better simply because this model has more free parameters, i.e. that the better fit does not necessarily imply that SMBHs grow in two phases.
    We can assess this by examining the parameter likelihoods for the 2P fits. 
    Figure \ref{fig:chi2} shows a corner plot for the fixed 2P fit (with $\alpha=1$). 
    The 2D panels quantify the degeneracies between pairs of parameters, while the 1D panels show the likelihoods for each of the five free parameters. 
    The likelihoods shown are profile likelihoods, 
    which correspond to maximum values of the 5D likelihood function as a function of the one or two parameters considered in each panel (i.e., the likelihood of the best-fit model with the one or two parameter values fixed and all others free). 
     First, we note that each of the free parameters is well constrained by the data, in that the profile likelihood has a well-defined peak within the range explored. 
     
     We can then ask whether the parameter values preferred by the 2P fit correspond to a ``physical'' two-phase model, in that the implied $\Mbh$-$\Mstar$ relation is consistent with the predictions of galaxy formation simulations that motivated our investigation (see Fig. \ref{fig:SMBH-Relation}). 
     This is a useful question because we constructed our 2P model to be sufficiently flexible that it can represent a wide range of $\Mbh$-$\Mstar$ relations, including a purely linear limit with no break, or a break of any magnitude at any stellar mass. 
     Using the results from the FIRE simulations shown in Figure \ref{fig:SMBH-Relation} as our primary reference point, the main characteristics of a physical two-phase model are a transition mass $\MstarCRIT \sim 5\times 10^{10}$ M$_{\odot}$ and a break factor $\gamma \gtrsim 10$. 
     The transition mass is roughly constant in the FIRE simulations (see also Byrne et al., in prep.). 
     The magnitude of the break factor is not robustly predicted by existing simulations because it is sensitive to the assumed ``seed'' mass for SMBHs, which determines the normalization of the $\Mbh$-$\Mstar$ relation at the low-mass end \citep[e.g.,][]{daniel+2017}. 
     We consider, however, that a two-phase scenario implies a break of at least one order of magnitude. 
     As Figure \ref{fig:chi2} shows, it is remarkable that the 2P fit to the QLF observations favors parameter values consistent with these simulation results 
     for a two-phase scenario, with a best-fit transition mass $\MstarCRIT \approx 10^{10.7} \solmass\approx 5\times10^{10} \solmass$ and a best-fit break factor $\gamma \approx 10^{1.8}\approx60$. 
     The best-fit high-mass normalization $\beta=M_{\rm BH}/M_{*}\sim0.5\%$ is furthermore similar to the value inferred more directly from BH mass measurements in massive galaxies \citep[e.g.,][]{MM13, kormendy&ho2013}, although we do not compare the exact values in detail because in our QLF analysis the $\beta$ normalization is degenerate with the assumed radiative efficiency.
    
    There is also some direct observational evidence for a break in the $\Mbh$-$\Mstar$ relation. 
    In a study that included AGN in dwarf galaxies (with BH masses inferred from broad lines), \cite{Reines2015} found that the low-mass galaxies in their sample had lower average $\Mbh$/$\Mstar$ than the high-mass galaxies with dynamical BH mass measurements (their Fig. 8). 
    Interestingly, the stellar mass and magnitude of the break in the $\Mbh$-$\Mstar$ relation from \cite{Reines2015} appear similar to the values favored by our 2P fits to the QLF (quoted above), albeit with large scatter. 
    We note, however, that in a study which measured stellar velocity dispersions $\sigma_{*}$ in eight active dwarf galaxies, \cite{Baldassare2020} found that the $\Mbh$-$\sigma_{*}$ relation for dwarfs is consistent with a power-law extrapolation from higher masses. 
    These results suggest that when testing models for scaling relations, it is important to consistently compare the models and observations for relations with respect to the same galaxy property (stellar mass, bulge mass, velocity dispersion, etc.). 
    As mentioned in the introduction, other observational studies have also reported evidence for changes in BH-galaxy scaling relations at low masses, for different cuts of the observational samples and different measures of host galaxy properties \citep{Graham+2013, Savorgnan+2016, Sahu+2019}.

We noted in the introduction that multiple different simulations predict two-phase SMBH growth qualitatively similar to what is shown for FIRE simulations in Figure \ref{fig:SMBH-Relation}. However, the different simulations do not agree in all quantitative details, so it is also interesting to consider how our best fits derived from the QLF data compare with the SMBH growth transitions found in other simulations. 
We focus here on comparing our results with the EAGLE simulations, for which there have been detailed studies of the SMBH growth transition  \citep[][]{bower+2017, delayedgrowth_EAGLE+2018}. 
Rather than stellar mass, \citet{delayedgrowth_EAGLE+2018} characterize the transition in terms of the properties of the dark matter halo, finding that the transition occurs in halos of a roughly constant virial temperature $T_{\rm vir}\approx 10^{5.6}$ K, corresponding to a critical halo mass that decreases with increasing redshift. 
To compare with \citet{delayedgrowth_EAGLE+2018}, we can infer the halo masses and virial temperatures corresponding to the transition stellar masses $\MstarCRIT$ favored by our fits. 
To do so, we use our ``evolving'' 2P fits in which the model parameters are allowed to evolve freely with redshift, so that we can infer how the preferred transition halo mass and virial temperature change with redshift. 
To infer halo mass from stellar mass, we use the median, redshift-dependent stellar mass-halo mass relation from UniverseMachine \citep[][]{behroozi+2019}. 
The virial temperature is then evaluated using standard relations for virialized halos \citep[][]{bryan+1998, Barkana_Loeb2001}.

\begin{figure}
            \centering
            \includegraphics[width=\linewidth, trim=0cm 0cm 0cm 0cm, clip=true]{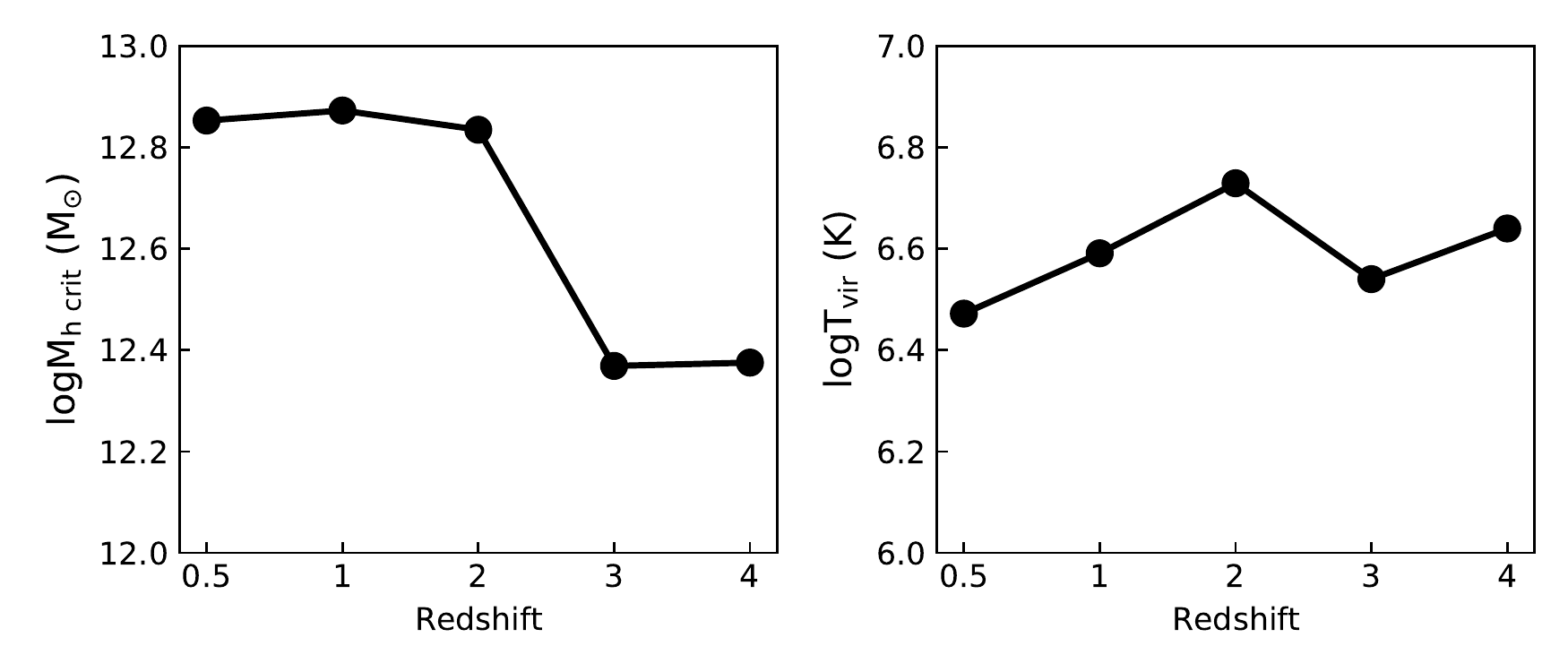}
            \caption{
            Halo mass and virial temperature corresponding to the redshift-dependent, best-fit transition stellar mass M$_{\rm *crit}$ for the ``evolving'' 2P fits (see the black dots in Fig. \ref{fig:bestfit_params}). 
            Halo masses are inferred assuming the median stellar mass-halo mass relation from UniverseMachine (Behroozi et al. 2019).
            }
            \label{fig:virial_temp}
        \end{figure}

The results for the implied transition halo mass and virial temperature vs. redshift are shown in Figure \ref{fig:virial_temp}. 
Interestingly, the 2P fits to the QLF data imply a transition halo virial temperature $T_{\rm vir}\sim 10^{6.6}$ K that is roughly constant over the redshift interval $z=0.5-4$ probed by our analysis, but a factor $\sim 10 \times$ higher than in the EAGLE simulations analyzed by \citet{delayedgrowth_EAGLE+2018}. 
This is a large systematic offset, corresponding to larger halo masses at the SMBH growth transition. 
We note, however, that the stellar mass-halo mass relation is relatively flat around the best-fit ``fixed'' $\MstarCRIT \approx 10^{10.7} \solmass$, so that a small change in stellar mass corresponds to large change in halo mass, according to the median relation. 
Moreover, in reality, there is scatter in the stellar mass-halo mass relation, as well as uncertainties in empirical determinations of this relation, such as the version in UniverseMachine that we have used here. 
To more robustly assess whether the QLF data may be consistent with a transition at a lower virial temperature similar to EAGLE,
it would be necessary to carefully model scatter and uncertainties in the stellar mass-halo mass relation, which is beyond the scope of the present work.

\section{Conclusions}\label{s:C}
    \subsection{Summary of Main Results}
	We use observations of the quasar luminosity function at $z\sim0.5-4$ to test a two-phase scenario for SMBH growth motivated by a number of recent galaxy formation simulations. 
	In this picture, SMBHs are undermassive relative to their host galaxies at low masses, or early times, leading to a $\Mbh$-$\Mstar$ scaling relation with a break and/or increased scatter below a certain mass. In the FIRE simulations, the break occurs at stellar mass $\MstarCRIT \sim 5\times10^{10} \solmass$ (e.g., Angl\'es-Alc\'azar et al. 2017b; Byrne et al., in prep.), though the exact threshold appears to depend on the simulation \citep[e.g.,][]{delayedgrowth_EAGLE+2018}. 
	We developed a flexible semi-analytic framework to predict the QLF implied for different two-phase model parameters, including a purely linear limit in which $\Mbh$ $\propto$ $\Mstar$ at all masses. 
	Our model incorporates a number of important physical constraints. 
	All models are anchored to the observed galaxy stellar mass function and star formation rates, as embodied by UniverseMachine \citep[][]{behroozi+2019}. 
    	The models are also self-consistent in that the BH growth rates are required to consistently connect the implied BH mass functions at all redshifts. 
	 Obscured BH growth is taken into account by comparing to the bolometric quasar luminosity function, which synthesizes observations from the IR to the X-rays \citep[][]{Shen2020}.
	
	The model fits statistically favor a two-phase scenario over a purely linear model with a difference in reduced $\chi^2$ values of $\Delta \chi^2_R=2.862$. 
	The two-phase model provides an excellent description of the observed QLF over the full luminosity range at $z=0.5$, where the observations are most complete and probe the largest luminosity range. 
	The linear model, on the other hand, fails to reproduce a well-defined QLF knee, 
	a problem which is especially severe at intermediate redshifts $z=2-3$, where a pronounced QLF knee is clearly implied by the data. 
	There is also some evidence that the data favor a two-phase scenario for physical reasons, rather than simply because the two-phase fits have more free parameters. 
	Namely, when the model parameters are free within wide ranges, the fits favor values for the transition mass and break factor in the $\Mbh$-$\Mstar$ relation that are very similar to the values predicted by simulations. 

    \subsection{Directions for Future Work}
    The main caveat to our conclusions is that, while our results indicate that a two-phase model is consistent with QLF observations, we cannot definitively rule out other scenarios. 
    For example, the models explored in this work assume that accretion rates follow a simple log-normal distribution. 
    It is possible that other forms of the accretion rate distribution would allow a linear model to better fit the observed QLF. 
    Other extensions of the model which could modify the results would be to include scatter in the scaling relations, allow for variable radiative efficiencies, or explicitly distinguish between total stellar mass and bulge mass. 
    It would also be interesting to explicitly include the effects of galaxy mergers. For example, \cite{delayedgrowth_EAGLE+2018} find that in EAGLE, rapid BH growth is often initiated by galaxy interactions, especially at low redshift, even though in their simulations a characteristic halo virial temperature remains a requirement for rapid growth.
    
    The model predictions should also be compared to other observations. 
    In \S \ref{s:D}, we mentioned measurements of BH-galaxy scaling relations extending into the dwarf regime \citep[see also][]{Reines2013, Mezcua2017}. 
    This kind of direct quantification of scaling relations across a wide range of galaxy masses provides stringent constraints on the models. 
    It would be valuable to more rigorously compare with such observations, accounting for selection effects which can affect scaling relations (e.g., galaxies selected purely based on stellar mass vs. actively accreting). 
    Focusing on active galaxies, it would be valuable to compare the models not only against the QLF (which integrates over the galaxy population at a given redshift) but also against more detailed measurements of accretion rates as a function of stellar mass and redshift (including summary statistics, such as active fractions, as well as full accretion rate distributions). 
    Such observations are already available \citep[e.g.,][]{Aird+2018II} and could help in breaking degeneracies. 
    As emphasized by \cite{Hickox+2014} and \cite{veale+2014}, comparing predictions for the distributions of host galaxy properties as a function of AGN luminosity (as opposed to AGN luminosity vs. host properties) can also distinguish between models that otherwise make similar predictions.  
    Finally, if massive BHs in low-mass galaxies depart substantially from standard scaling relations, this would have important implications for expected BH merger rates in dwarf galaxies, and in the early universe in particular. 
    The effects of this may be detectable by future spaced-based gravitational wave detectors \citep[][]{Bailes2021_roadmap}.

\section*{Acknowledgements}
    We thank the referee, Stuart McAlpine, for a very constructive review. 
    MTT thanks Aaron Geller for hosting the 2018 Northwestern CIERA REU students and for welcoming her back in the summer of 2019. 
    MTT also thanks Kim-Vy Tran and Louis E. Strigari for mentoring her Texas A\&M University undergraduate thesis project based on this work.
    We thank Alex Gurvich for help with code, Alex Richings for data on bulge fractions, Jacob Shen and Phil Hopkins for collaboration on the quasar luminosity function, Philip Arevalo for advice on effective statistical presentations, and Tjitske Starkenburg, Michael Grudic, Jonathan Stern, Lindsey Byrne, and Zachary Hafen for advice during this project. 
    This material is based upon work supported by NSF Grant No. AST-1757792. SW is supported by an NSF Astronomy and Astrophysics Postdoctoral Fellowship under award AST2001905. CAFG was supported by NSF through grants AST-1715216, AST-2108230, and CAREER award AST-1652522; by NASA through grant 17-ATP17-0067; by STScI through grant HST-AR-16124.001-A; and by the Research Corporation for Science Advancement through a Cottrell Scholar Award and a Scialog Award.
    This work was performed in part at Aspen Center for Physics, which is supported by National Science Foundation grant PHY-1607611. DAA was supported in part by NSF grants AST-2009687 and AST-2108944, and by the Flatiron Institute, which is supported by the Simons Foundation.

\section*{Data Availability}
The methods described in the paper should be sufficient to reproduce most results. 
Additional data can be obtained from the corresponding author on reasonable request.

\bibliographystyle{mnras}
\bibliography{mybib} 

\begin{thebibliography}{}
\makeatletter
\relax
\def\mn@urlcharsother{\let\do\@makeother \do\$\do\&\do\#\do\^\do\_\do\%\do\~}
\def\mn@doi{\begingroup\mn@urlcharsother \@ifnextchar [ {\mn@doi@}
  {\mn@doi@[]}}
\def\mn@doi@[#1]#2{\def\@tempa{#1}\ifx\@tempa\@empty \href
  {http://dx.doi.org/#2} {doi:#2}\else \href {http://dx.doi.org/#2} {#1}\fi
  \endgroup}
\def\mn@eprint#1#2{\mn@eprint@#1:#2::\@nil}
\def\mn@eprint@arXiv#1{\href {http://arxiv.org/abs/#1} {{\tt arXiv:#1}}}
\def\mn@eprint@dblp#1{\href {http://dblp.uni-trier.de/rec/bibtex/#1.xml}
  {dblp:#1}}
\def\mn@eprint@#1:#2:#3:#4\@nil{\def\@tempa {#1}\def\@tempb {#2}\def\@tempc
  {#3}\ifx \@tempc \@empty \let \@tempc \@tempb \let \@tempb \@tempa \fi \ifx
  \@tempb \@empty \def\@tempb {arXiv}\fi \@ifundefined
  {mn@eprint@\@tempb}{\@tempb:\@tempc}{\expandafter \expandafter \csname
  mn@eprint@\@tempb\endcsname \expandafter{\@tempc}}}

\bibitem[\protect\citeauthoryear{{Abramowicz} \& {Fragile}}{{Abramowicz} \&
  {Fragile}}{2013}]{Abramowicz+2013}
{Abramowicz} M.~A.,  {Fragile} P.~C.,  2013, \mn@doi [Living Reviews in
  Relativity] {10.12942/lrr-2013-1}, \href
  {https://ui.adsabs.harvard.edu/abs/2013LRR....16....1A} {16, 1}

\bibitem[\protect\citeauthoryear{{Aird}, {Coil}  \& {Georgakakis}}{{Aird}
  et~al.}{2018}]{Aird+2018II}
{Aird} J.,  {Coil} A.~L.,   {Georgakakis} A.,  2018, \mn@doi [\mnras]
  {10.1093/mnras/stx2700}, \href
  {https://ui.adsabs.harvard.edu/abs/2018MNRAS.474.1225A} {474, 1225}

\bibitem[\protect\citeauthoryear{{Angl{\'e}s-Alc{\'a}zar},
  {Faucher-Gigu{\`e}re}, {Kere{\v{s}}}, {Hopkins}, {Quataert}  \&
  {Murray}}{{Angl{\'e}s-Alc{\'a}zar} et~al.}{2017a}]{AA17_cycle}
{Angl{\'e}s-Alc{\'a}zar} D.,  {Faucher-Gigu{\`e}re} C.-A.,  {Kere{\v{s}}} D.,
  {Hopkins} P.~F.,  {Quataert} E.,   {Murray} N.,  2017a, \mn@doi [\mnras]
  {10.1093/mnras/stx1517}, \href
  {https://ui.adsabs.harvard.edu/abs/2017MNRAS.470.4698A} {470, 4698}

\bibitem[\protect\citeauthoryear{{Angl{\'e}s-Alc{\'a}zar},
  {Faucher-Gigu{\`e}re}, {Quataert}, {Hopkins}, {Feldmann}, {Torrey}, {Wetzel}
  \& {Kere{\v s}}}{{Angl{\'e}s-Alc{\'a}zar} et~al.}{2017b}]{daniel+2017}
{Angl{\'e}s-Alc{\'a}zar} D.,  {Faucher-Gigu{\`e}re} C.-A.,  {Quataert} E.,
  {Hopkins} P.~F.,  {Feldmann} R.,  {Torrey} P.,  {Wetzel} A.,   {Kere{\v s}}
  D.,  2017b, \mn@doi [\mnras] {10.1093/mnrasl/slx161}, \href
  {http://adsabs.harvard.edu/abs/2017MNRAS.472L.109A} {472, L109}

\bibitem[\protect\citeauthoryear{{Angl{\'e}s-Alc{\'a}zar}
  et~al.,}{{Angl{\'e}s-Alc{\'a}zar} et~al.}{2021}]{AA21_hyper}
{Angl{\'e}s-Alc{\'a}zar} D.,  et~al., 2021, \mn@doi [\apj]
  {10.3847/1538-4357/ac09e8}, \href
  {https://ui.adsabs.harvard.edu/abs/2021ApJ...917...53A} {917, 53}

\bibitem[\protect\citeauthoryear{{Bailes} et~al.,}{{Bailes}
  et~al.}{2021}]{Bailes2021_roadmap}
{Bailes} M.,  et~al., 2021, \mn@doi [Nature Reviews Physics]
  {10.1038/s42254-021-00303-8}, \href
  {https://ui.adsabs.harvard.edu/abs/2021NatRP...3..344B} {3, 344}

\bibitem[\protect\citeauthoryear{{Baldassare}, {Dickey}, {Geha}  \&
  {Reines}}{{Baldassare} et~al.}{2020}]{Baldassare2020}
{Baldassare} V.~F.,  {Dickey} C.,  {Geha} M.,   {Reines} A.~E.,  2020, \mn@doi
  [\apjl] {10.3847/2041-8213/aba0c1}, \href
  {https://ui.adsabs.harvard.edu/abs/2020ApJ...898L...3B} {898, L3}

\bibitem[\protect\citeauthoryear{{Barkana} \& {Loeb}}{{Barkana} \&
  {Loeb}}{2001}]{Barkana_Loeb2001}
{Barkana} R.,  {Loeb} A.,  2001, \mn@doi [\physrep]
  {10.1016/S0370-1573(01)00019-9}, \href
  {https://ui.adsabs.harvard.edu/abs/2001PhR...349..125B} {349, 125}

\bibitem[\protect\citeauthoryear{{Behroozi}, {Wechsler}, {Hearin}  \&
  {Conroy}}{{Behroozi} et~al.}{2019}]{behroozi+2019}
{Behroozi} P.,  {Wechsler} R.~H.,  {Hearin} A.~P.,   {Conroy} C.,  2019,
  \mn@doi [\mnras] {10.1093/mnras/stz1182}, \href
  {https://ui.adsabs.harvard.edu/abs/2019MNRAS.488.3143B} {488, 3143}

\bibitem[\protect\citeauthoryear{Benson, Danovic, Frenk  \& Sharples}{Benson
  et~al.}{2007}]{benson+2007}
Benson A.~J.,  Danovic D.,  Frenk C.~S.,   Sharples R.,  2007, \mn@doi [Monthly
  Notices of the Royal Astronomical Society]
  {10.1111/j.1365-2966.2007.11923.x}, 379, 841

\bibitem[\protect\citeauthoryear{Bonoli, Mayer, Kazantzidis, Madau, Bellovary
  \& Governato}{Bonoli et~al.}{2016}]{bonoli+2016}
Bonoli S.,  Mayer L.,  Kazantzidis S.,  Madau P.,  Bellovary J.,   Governato
  F.,  2016, \mn@doi [Monthly Notices of the Royal Astronomical Society]
  {10.1093/mnras/stw694}, 459, 2603

\bibitem[\protect\citeauthoryear{{Bower}, {Schaye}, {Frenk}, {Theuns},
  {Schaller}, {Crain}  \& {McAlpine}}{{Bower} et~al.}{2017}]{bower+2017}
{Bower} R.~G.,  {Schaye} J.,  {Frenk} C.~S.,  {Theuns} T.,  {Schaller} M.,
  {Crain} R.~A.,   {McAlpine} S.,  2017, \mn@doi [\mnras]
  {10.1093/mnras/stw2735}, \href
  {https://ui.adsabs.harvard.edu/abs/2017MNRAS.465...32B} {465, 32}

\bibitem[\protect\citeauthoryear{{Bryan} \& {Norman}}{{Bryan} \&
  {Norman}}{1998}]{bryan+1998}
{Bryan} G.~L.,  {Norman} M.~L.,  1998, \mn@doi [\apj] {10.1086/305262}, \href
  {https://ui.adsabs.harvard.edu/abs/1998ApJ...495...80B} {495, 80}

\bibitem[\protect\citeauthoryear{{Chen} et~al.,}{{Chen}
  et~al.}{2020}]{Chen2020}
{Chen} Z.,  et~al., 2020, \mn@doi [\apj] {10.3847/1538-4357/ab9633}, \href
  {https://ui.adsabs.harvard.edu/abs/2020ApJ...897..102C} {897, 102}

\bibitem[\protect\citeauthoryear{{Conroy} \& {White}}{{Conroy} \&
  {White}}{2013}]{conroy+white2013}
{Conroy} C.,  {White} M.,  2013, \mn@doi [\apj] {10.1088/0004-637X/762/2/70},
  \href {https://ui.adsabs.harvard.edu/abs/2013ApJ...762...70C} {762, 70}

\bibitem[\protect\citeauthoryear{{Dubois}, {Volonteri}, {Silk}, {Devriendt},
  {Slyz}  \& {Teyssier}}{{Dubois} et~al.}{2015}]{dubois+2015}
{Dubois} Y.,  {Volonteri} M.,  {Silk} J.,  {Devriendt} J.,  {Slyz} A.,
  {Teyssier} R.,  2015, \mn@doi [\mnras] {10.1093/mnras/stv1416}, \href
  {https://ui.adsabs.harvard.edu/abs/2015MNRAS.452.1502D} {452, 1502}

\bibitem[\protect\citeauthoryear{{Faber} et~al.,}{{Faber}
  et~al.}{2007}]{Faber2007}
{Faber} S.~M.,  et~al., 2007, \mn@doi [\apj] {10.1086/519294}, \href
  {https://ui.adsabs.harvard.edu/abs/2007ApJ...665..265F} {665, 265}

\bibitem[\protect\citeauthoryear{{Faucher-Gigu{\`e}re}}{{Faucher-Gigu{\`e}re}}{2018}]{FG18_NA}
{Faucher-Gigu{\`e}re} C.-A.,  2018, \mn@doi [Nature Astronomy]
  {10.1038/s41550-018-0427-y}, \href
  {https://ui.adsabs.harvard.edu/abs/2018NatAs...2..368F} {2, 368}

\bibitem[\protect\citeauthoryear{{Ferrarese} \& {Merritt}}{{Ferrarese} \&
  {Merritt}}{2000}]{Ferrarese2000}
{Ferrarese} L.,  {Merritt} D.,  2000, \mn@doi [\apjl] {10.1086/312838}, \href
  {https://ui.adsabs.harvard.edu/abs/2000ApJ...539L...9F} {539, L9}

\bibitem[\protect\citeauthoryear{{Gebhardt} et~al.,}{{Gebhardt}
  et~al.}{2000}]{Gebhardt2000}
{Gebhardt} K.,  et~al., 2000, \mn@doi [\apjl] {10.1086/312840}, \href
  {https://ui.adsabs.harvard.edu/abs/2000ApJ...539L..13G} {539, L13}

\bibitem[\protect\citeauthoryear{{Graham} \& {Scott}}{{Graham} \&
  {Scott}}{2013}]{Graham+2013}
{Graham} A.~W.,  {Scott} N.,  2013, \mn@doi [\apj]
  {10.1088/0004-637X/764/2/151}, \href
  {https://ui.adsabs.harvard.edu/abs/2013ApJ...764..151G} {764, 151}

\bibitem[\protect\citeauthoryear{{Habouzit}, {Volonteri}  \&
  {Dubois}}{{Habouzit} et~al.}{2017}]{habouzit+2017}
{Habouzit} M.,  {Volonteri} M.,   {Dubois} Y.,  2017, \mn@doi [\mnras]
  {10.1093/mnras/stx666}, \href
  {https://ui.adsabs.harvard.edu/abs/2017MNRAS.468.3935H} {468, 3935}

\bibitem[\protect\citeauthoryear{{H{\"a}ring} \& {Rix}}{{H{\"a}ring} \&
  {Rix}}{2004}]{haring+2004}
{H{\"a}ring} N.,  {Rix} H.-W.,  2004, \mn@doi [\apjl] {10.1086/383567}, \href
  {https://ui.adsabs.harvard.edu/abs/2004ApJ...604L..89H} {604, L89}

\bibitem[\protect\citeauthoryear{{Hickox} \& {Alexander}}{{Hickox} \&
  {Alexander}}{2018}]{HA2018}
{Hickox} R.~C.,  {Alexander} D.~M.,  2018, \mn@doi [\araa]
  {10.1146/annurev-astro-081817-051803}, \href
  {https://ui.adsabs.harvard.edu/abs/2018ARA&A..56..625H} {56, 625}

\bibitem[\protect\citeauthoryear{{Hickox}, {Mullaney}, {Alexander}, {Chen},
  {Civano}, {Goulding}  \& {Hainline}}{{Hickox} et~al.}{2014}]{Hickox+2014}
{Hickox} R.~C.,  {Mullaney} J.~R.,  {Alexander} D.~M.,  {Chen} C.-T.~J.,
  {Civano} F.~M.,  {Goulding} A.~D.,   {Hainline} K.~N.,  2014, \mn@doi [\apj]
  {10.1088/0004-637X/782/1/9}, \href
  {https://ui.adsabs.harvard.edu/abs/2014ApJ...782....9H} {782, 9}

\bibitem[\protect\citeauthoryear{{Hopkins}, {Richards}  \&
  {Hernquist}}{{Hopkins} et~al.}{2007}]{Hopkins2007}
{Hopkins} P.~F.,  {Richards} G.~T.,   {Hernquist} L.,  2007, \mn@doi [\apj]
  {10.1086/509629}, \href
  {https://ui.adsabs.harvard.edu/abs/2007ApJ...654..731H} {654, 731}

\bibitem[\protect\citeauthoryear{{Hopkins}, {Cox}, {Kere{\v{s}}}  \&
  {Hernquist}}{{Hopkins} et~al.}{2008}]{Hopkins2008_red_ellipticals}
{Hopkins} P.~F.,  {Cox} T.~J.,  {Kere{\v{s}}} D.,   {Hernquist} L.,  2008,
  \mn@doi [\apjs] {10.1086/524363}, \href
  {https://ui.adsabs.harvard.edu/abs/2008ApJS..175..390H} {175, 390}

\bibitem[\protect\citeauthoryear{{Hopkins}, {Kere{\v{s}}}, {O{\~n}orbe},
  {Faucher-Gigu{\`e}re}, {Quataert}, {Murray}  \& {Bullock}}{{Hopkins}
  et~al.}{2014}]{hopkins+2014}
{Hopkins} P.~F.,  {Kere{\v{s}}} D.,  {O{\~n}orbe} J.,  {Faucher-Gigu{\`e}re}
  C.-A.,  {Quataert} E.,  {Murray} N.,   {Bullock} J.~S.,  2014, \mn@doi
  [\mnras] {10.1093/mnras/stu1738}, \href
  {https://ui.adsabs.harvard.edu/abs/2014MNRAS.445..581H} {445, 581}

\bibitem[\protect\citeauthoryear{{Hopkins}, {Torrey}, {Faucher-Gigu{\`e}re},
  {Quataert}  \& {Murray}}{{Hopkins} et~al.}{2016}]{Hopkins16_concert}
{Hopkins} P.~F.,  {Torrey} P.,  {Faucher-Gigu{\`e}re} C.-A.,  {Quataert} E.,
  {Murray} N.,  2016, \mn@doi [\mnras] {10.1093/mnras/stw289}, \href
  {https://ui.adsabs.harvard.edu/abs/2016MNRAS.458..816H} {458, 816}

\bibitem[\protect\citeauthoryear{{Hopkins} et~al.,}{{Hopkins}
  et~al.}{2018}]{hopkins+2018}
{Hopkins} P.~F.,  et~al., 2018, \mn@doi [\mnras] {10.1093/mnras/sty1690}, \href
  {https://ui.adsabs.harvard.edu/abs/2018MNRAS.480..800H} {480, 800}

\bibitem[\protect\citeauthoryear{{Kormendy} \& {Ho}}{{Kormendy} \&
  {Ho}}{2013}]{kormendy&ho2013}
{Kormendy} J.,  {Ho} L.~C.,  2013, \mn@doi [\araa]
  {10.1146/annurev-astro-082708-101811}, \href
  {https://ui.adsabs.harvard.edu/abs/2013ARA%26A..51..511K} {51, 511}

\bibitem[\protect\citeauthoryear{{Lapiner}, {Dekel}  \& {Dubois}}{{Lapiner}
  et~al.}{2021}]{Lapiner2021}
{Lapiner} S.,  {Dekel} A.,   {Dubois} Y.,  2021, \mn@doi [\mnras]
  {10.1093/mnras/stab1205}, \href
  {https://ui.adsabs.harvard.edu/abs/2021MNRAS.505..172L} {505, 172}

\bibitem[\protect\citeauthoryear{{L{\"a}sker}, {Greene}, {Seth}, {van de Ven},
  {Braatz}, {Henkel}  \& {Lo}}{{L{\"a}sker} et~al.}{2016}]{Lasker2016}
{L{\"a}sker} R.,  {Greene} J.~E.,  {Seth} A.,  {van de Ven} G.,  {Braatz}
  J.~A.,  {Henkel} C.,   {Lo} K.~Y.,  2016, \mn@doi [\apj]
  {10.3847/0004-637X/825/1/3}, \href
  {https://ui.adsabs.harvard.edu/abs/2016ApJ...825....3L} {825, 3}

\bibitem[\protect\citeauthoryear{{Leitner} \& {Kravtsov}}{{Leitner} \&
  {Kravtsov}}{2011}]{Leitner2011}
{Leitner} S.~N.,  {Kravtsov} A.~V.,  2011, \mn@doi [\apj]
  {10.1088/0004-637X/734/1/48}, \href
  {https://ui.adsabs.harvard.edu/abs/2011ApJ...734...48L} {734, 48}

\bibitem[\protect\citeauthoryear{{Liu}, {Veilleux}, {Canalizo}, {Rupke},
  {Manzano-King}, {Bohn}  \& {U}}{{Liu} et~al.}{2020}]{Liu2020}
{Liu} W.,  {Veilleux} S.,  {Canalizo} G.,  {Rupke} D. S.~N.,  {Manzano-King}
  C.~M.,  {Bohn} T.,   {U} V.,  2020, \mn@doi [\apj]
  {10.3847/1538-4357/abc269}, \href
  {https://ui.adsabs.harvard.edu/abs/2020ApJ...905..166L} {905, 166}

\bibitem[\protect\citeauthoryear{{Ma}, {Hopkins}, {Ma},
  {Angl{\'e}s-Alc{\'a}zar}, {Faucher-Gigu{\`e}re}  \& {Kelley}}{{Ma}
  et~al.}{2021}]{Ma2021_sinking}
{Ma} L.,  {Hopkins} P.~F.,  {Ma} X.,  {Angl{\'e}s-Alc{\'a}zar} D.,
  {Faucher-Gigu{\`e}re} C.-A.,   {Kelley} L.~Z.,  2021, arXiv e-prints, \href
  {https://ui.adsabs.harvard.edu/abs/2021arXiv210102727M} {p. arXiv:2101.02727}

\bibitem[\protect\citeauthoryear{{Magorrian} et~al.,}{{Magorrian}
  et~al.}{1998}]{Magorrian1998}
{Magorrian} J.,  et~al., 1998, \mn@doi [\aj] {10.1086/300353}, \href
  {https://ui.adsabs.harvard.edu/abs/1998AJ....115.2285M} {115, 2285}

\bibitem[\protect\citeauthoryear{{Manzano-King}, {Canalizo}  \&
  {Sales}}{{Manzano-King} et~al.}{2019}]{ManzanoKing2019}
{Manzano-King} C.~M.,  {Canalizo} G.,   {Sales} L.~V.,  2019, \mn@doi [\apj]
  {10.3847/1538-4357/ab4197}, \href
  {https://ui.adsabs.harvard.edu/abs/2019ApJ...884...54M} {884, 54}

\bibitem[\protect\citeauthoryear{{Marconi} \& {Hunt}}{{Marconi} \&
  {Hunt}}{2003}]{marconi+2003}
{Marconi} A.,  {Hunt} L.~K.,  2003, \mn@doi [\apjl] {10.1086/375804}, \href
  {https://ui.adsabs.harvard.edu/abs/2003ApJ...589L..21M} {589, L21}

\bibitem[\protect\citeauthoryear{{McAlpine}, {Bower}, {Rosario}, {Crain},
  {Schaye}  \& {Theuns}}{{McAlpine} et~al.}{2018}]{delayedgrowth_EAGLE+2018}
{McAlpine} S.,  {Bower} R.~G.,  {Rosario} D.~J.,  {Crain} R.~A.,  {Schaye} J.,
   {Theuns} T.,  2018, \mn@doi [\mnras] {10.1093/mnras/sty2489}, \href
  {https://ui.adsabs.harvard.edu/abs/2018MNRAS.481.3118M} {481, 3118}

\bibitem[\protect\citeauthoryear{{McConnell} \& {Ma}}{{McConnell} \&
  {Ma}}{2013}]{MM13}
{McConnell} N.~J.,  {Ma} C.-P.,  2013, \mn@doi [\apj]
  {10.1088/0004-637X/764/2/184}, \href
  {https://ui.adsabs.harvard.edu/abs/2013ApJ...764..184M} {764, 184}

\bibitem[\protect\citeauthoryear{{Merloni} et~al.,}{{Merloni}
  et~al.}{2010}]{Merloni2010}
{Merloni} A.,  et~al., 2010, \mn@doi [\apj] {10.1088/0004-637X/708/1/137},
  \href {https://ui.adsabs.harvard.edu/abs/2010ApJ...708..137M} {708, 137}

\bibitem[\protect\citeauthoryear{{Mezcua}}{{Mezcua}}{2017}]{Mezcua2017}
{Mezcua} M.,  2017, \mn@doi [International Journal of Modern Physics D]
  {10.1142/S021827181730021X}, \href
  {https://ui.adsabs.harvard.edu/abs/2017IJMPD..2630021M} {26, 1730021}

\bibitem[\protect\citeauthoryear{{Naab} \& {Ostriker}}{{Naab} \&
  {Ostriker}}{2017}]{NO2017_ARAA}
{Naab} T.,  {Ostriker} J.~P.,  2017, \mn@doi [\araa]
  {10.1146/annurev-astro-081913-040019}, \href
  {https://ui.adsabs.harvard.edu/abs/2017ARA&A..55...59N} {55, 59}

\bibitem[\protect\citeauthoryear{{Nguyen} et~al.,}{{Nguyen}
  et~al.}{2019}]{Nguyen2019}
{Nguyen} D.~D.,  et~al., 2019, \mn@doi [\apj] {10.3847/1538-4357/aafe7a}, \href
  {https://ui.adsabs.harvard.edu/abs/2019ApJ...872..104N} {872, 104}

\bibitem[\protect\citeauthoryear{{Prieto}, {Escala}, {Volonteri}  \&
  {Dubois}}{{Prieto} et~al.}{2017}]{prieto+2017}
{Prieto} J.,  {Escala} A.,  {Volonteri} M.,   {Dubois} Y.,  2017, \mn@doi
  [\apj] {10.3847/1538-4357/aa5be5}, \href
  {https://ui.adsabs.harvard.edu/abs/2017ApJ...836..216P} {836, 216}

\bibitem[\protect\citeauthoryear{{Reines} \& {Volonteri}}{{Reines} \&
  {Volonteri}}{2015}]{Reines2015}
{Reines} A.~E.,  {Volonteri} M.,  2015, \mn@doi [\apj]
  {10.1088/0004-637X/813/2/82}, \href
  {https://ui.adsabs.harvard.edu/abs/2015ApJ...813...82R} {813, 82}

\bibitem[\protect\citeauthoryear{{Reines}, {Greene}  \& {Geha}}{{Reines}
  et~al.}{2013}]{Reines2013}
{Reines} A.~E.,  {Greene} J.~E.,   {Geha} M.,  2013, \mn@doi [\apj]
  {10.1088/0004-637X/775/2/116}, \href
  {https://ui.adsabs.harvard.edu/abs/2013ApJ...775..116R} {775, 116}

\bibitem[\protect\citeauthoryear{Sahu, Graham  \& Davis}{Sahu
  et~al.}{2019}]{Sahu+2019}
Sahu N.,  Graham A.~W.,   Davis B.~L.,  2019, \mn@doi [The Astrophysical
  Journal] {10.3847/1538-4357/ab0f32}, 876, 155

\bibitem[\protect\citeauthoryear{{Savorgnan}}{{Savorgnan}}{2016}]{Savorgnan+2016}
{Savorgnan} G. A.~D.,  2016, \mn@doi [\apj] {10.3847/0004-637X/821/2/88}, \href
  {https://ui.adsabs.harvard.edu/abs/2016ApJ...821...88S} {821, 88}

\bibitem[\protect\citeauthoryear{{Schutte}, {Reines}  \& {Greene}}{{Schutte}
  et~al.}{2019}]{Schutte2019}
{Schutte} Z.,  {Reines} A.~E.,   {Greene} J.~E.,  2019, \mn@doi [\apj]
  {10.3847/1538-4357/ab35dd}, \href
  {https://ui.adsabs.harvard.edu/abs/2019ApJ...887..245S} {887, 245}

\bibitem[\protect\citeauthoryear{{Shen} et~al.,}{{Shen}
  et~al.}{2015}]{Shen2015}
{Shen} Y.,  et~al., 2015, \mn@doi [\apj] {10.1088/0004-637X/805/2/96}, \href
  {https://ui.adsabs.harvard.edu/abs/2015ApJ...805...96S} {805, 96}

\bibitem[\protect\citeauthoryear{{Shen}, {Hopkins}, {Faucher-Gigu{\`e}re},
  {Alexander}, {Richards}, {Ross}  \& {Hickox}}{{Shen} et~al.}{2020}]{Shen2020}
{Shen} X.,  {Hopkins} P.~F.,  {Faucher-Gigu{\`e}re} C.-A.,  {Alexander} D.~M.,
  {Richards} G.~T.,  {Ross} N.~P.,   {Hickox} R.~C.,  2020, \mn@doi [\mnras]
  {10.1093/mnras/staa1381}, \href
  {https://ui.adsabs.harvard.edu/abs/2020MNRAS.495.3252S} {495, 3252}

\bibitem[\protect\citeauthoryear{{Small} \& {Blandford}}{{Small} \&
  {Blandford}}{1992}]{SB1992}
{Small} T.~A.,  {Blandford} R.~D.,  1992, \mn@doi [\mnras]
  {10.1093/mnras/259.4.725}, \href
  {https://ui.adsabs.harvard.edu/abs/1992MNRAS.259..725S} {259, 725}

\bibitem[\protect\citeauthoryear{{Soltan}}{{Soltan}}{1982}]{Soltan_1982}
{Soltan} A.,  1982, \mn@doi [\mnras] {10.1093/mnras/200.1.115}, \href
  {https://ui.adsabs.harvard.edu/abs/1982MNRAS.200..115S} {200, 115}

\bibitem[\protect\citeauthoryear{{Somerville} \& {Dav{\'e}}}{{Somerville} \&
  {Dav{\'e}}}{2015}]{SD2015_ARAA}
{Somerville} R.~S.,  {Dav{\'e}} R.,  2015, \mn@doi [\araa]
  {10.1146/annurev-astro-082812-140951}, \href
  {https://ui.adsabs.harvard.edu/abs/2015ARA&A..53...51S} {53, 51}

\bibitem[\protect\citeauthoryear{{Stern} et~al.,}{{Stern}
  et~al.}{2020}]{Stern+2020}
{Stern} J.,  et~al., 2020, arXiv e-prints, \href
  {https://ui.adsabs.harvard.edu/abs/2020arXiv200613976S} {p. arXiv:2006.13976}

\bibitem[\protect\citeauthoryear{{Tremaine} et~al.,}{{Tremaine}
  et~al.}{2002}]{Tremaine2002}
{Tremaine} S.,  et~al., 2002, \mn@doi [\apj] {10.1086/341002}, \href
  {https://ui.adsabs.harvard.edu/abs/2002ApJ...574..740T} {574, 740}

\bibitem[\protect\citeauthoryear{{Treu}, {Woo}, {Malkan}  \&
  {Blandford}}{{Treu} et~al.}{2007}]{Treu2007}
{Treu} T.,  {Woo} J.-H.,  {Malkan} M.~A.,   {Blandford} R.~D.,  2007, \mn@doi
  [\apj] {10.1086/520633}, \href
  {https://ui.adsabs.harvard.edu/abs/2007ApJ...667..117T} {667, 117}

\bibitem[\protect\citeauthoryear{Veale, White  \& Conroy}{Veale
  et~al.}{2014}]{veale+2014}
Veale M.,  White M.,   Conroy C.,  2014, \mn@doi [Monthly Notices of the Royal
  Astronomical Society] {10.1093/mnras/stu1821}, 445, 1144

\bibitem[\protect\citeauthoryear{{Vogelsberger}, {Marinacci}, {Torrey}  \&
  {Puchwein}}{{Vogelsberger} et~al.}{2020}]{Vogelsberger2020_NatRP}
{Vogelsberger} M.,  {Marinacci} F.,  {Torrey} P.,   {Puchwein} E.,  2020,
  \mn@doi [Nature Reviews Physics] {10.1038/s42254-019-0127-2}, \href
  {https://ui.adsabs.harvard.edu/abs/2020NatRP...2...42V} {2, 42}

\bibitem[\protect\citeauthoryear{{Yu} \& {Tremaine}}{{Yu} \&
  {Tremaine}}{2002}]{YT2002}
{Yu} Q.,  {Tremaine} S.,  2002, \mn@doi [\mnras]
  {10.1046/j.1365-8711.2002.05532.x}, \href
  {https://ui.adsabs.harvard.edu/abs/2002MNRAS.335..965Y} {335, 965}

\bibitem[\protect\citeauthoryear{{Zhang}, {Behroozi}, {Volonteri}, {Silk},
  {Fan}, {Hopkins}, {Yang}  \& {Aird}}{{Zhang}
  et~al.}{2021}]{Zhang2021_Trinity}
{Zhang} H.,  {Behroozi} P.,  {Volonteri} M.,  {Silk} J.,  {Fan} X.,  {Hopkins}
  P.~F.,  {Yang} J.,   {Aird} J.,  2021, arXiv e-prints, \href
  {https://ui.adsabs.harvard.edu/abs/2021arXiv210510474Z} {p. arXiv:2105.10474}

\bibitem[\protect\citeauthoryear{{{\c{C}}atmabacak}, {Feldmann},
  {Angl{\'e}s-Alc{\'a}zar}, {Faucher-Gigu{\`e}re}, {Hopkins}  \&
  {Kere{\v{s}}}}{{{\c{C}}atmabacak} et~al.}{2020}]{Catmabacak2021}
{{\c{C}}atmabacak} O.,  {Feldmann} R.,  {Angl{\'e}s-Alc{\'a}zar} D.,
  {Faucher-Gigu{\`e}re} C.-A.,  {Hopkins} P.~F.,   {Kere{\v{s}}} D.,  2020,
  arXiv e-prints, \href {https://ui.adsabs.harvard.edu/abs/2020arXiv200712185C}
  {p. arXiv:2007.12185}

\makeatother
\end{thebibliography}

\bsp	
\label{lastpage}
\end{document}